\documentclass[amsmath,amssymb,aps,twocolumn,prl,longbibliography]{revtex4-2}
\usepackage{amssymb}
\usepackage{graphicx}
\usepackage{dcolumn}
\usepackage{bm}
\usepackage{amsmath}
\usepackage{subfigure}
\usepackage{float}
\usepackage{color}
\usepackage[colorlinks=true,citecolor=blue]{hyperref}

\begin{document}

\preprint{}

\title{Single photonic qutrit in a collective Rydberg polariton}


\author{Yuechun Jiao$^{1,2}$}
\author{Oliver D. W. Hughes$^{1}$}
 \author{Max Z. Festenstein$^{1}$}
\author{Zhengyang Bai$^{3}$}
\thanks{zhybai@lps.ecnu.edu.cn}
\author{Jianming Zhao$^{2}$}
\author{Weibin Li$^{4}$}
\author{Kevin J. Weatherill$^{1}$}
\author{C. Stuart Adams$^{1}$}
\thanks{c.s.adams@durham.ac.uk}

\affiliation{
$^{1}$Joint Quantum Centre (Durham-Newcastle), Department of Physics, Durham University, Durham, DH1 3LE, United Kingdom\\
$^{2}$State Key Laboratory of Quantum Optics and Quantum Optics Devices, Institute of Laser Spectroscopy, Shanxi University, Taiyuan 030006, China\\
$^{3}$State Key Laboratory of Precision Spectroscopy, East China Normal University, Shanghai 200062, China\\
$^{4}$School of Physics and Astronomy and Centre for the Mathematics and Theoretical Physics of Quantum Non-equilibrium Systems, University of Nottingham, Nottingham NG7 2RD, United Kingdom}

\date{\today}

\begin{abstract}
We report on the coherent creation, control and read-out of a single photonic qutrit in a Rydberg ensemble. In each measurement, an optical photon is stored as a Rydberg polariton through electromagnetically induced transparency. Employing two microwave fields, the polariton is driven into an arbitrary superposition of three collective states, each encoded in a Rydberg state. The collective state is mapped into a photonic time-bin qutrit with the microwave field and read out sequentially. The complete sequence, including preparation, control, and read-out, is less than 1.8~$\mu$s, which mitigates decoherence significantly. We measure the coherence of the qutrit with non-destructive Ramsey interferometry, which is preferable for quantum information processing, and find good quantitative agreement with the theoretical model. The ability to write, process and read out the single photonic qutrit on microsecond time scales with microwave coupled Rydberg states demonstrates the coherent connectivity among the high Hilbert space of the qutrit.Our study is an important step in exploring qutrit based quantum information processes and quantum simulation of topological physics with microwave coupled Rydberg atom ensembles.

\end{abstract}

\maketitle

\textit {Introduction---} 
Qutrits and more generally qudits, offer higher dimensional computational space, which can significantly enhance performance and efficiency in computation and communication~\cite{blok2021,mackeprang2023,wang2020Qudits,Robust_2021_PRL} through highly efficient algorithms \cite{Fedorov2012,Gokhale2019}, increase quantum security \cite{Vaziri2003,Lanyon2008,Luo2019}, test fundamental quantum mechanics~\cite{collins2002BellInequalitiesArbitrarily} and explore topological phases~\cite{liu2021}, respectively. Higher dimensional qudit space enables quantum solvers to address new classes of optimization problems such as graph colouring~\cite{acin2024}, and 
herald measurement of which-path information~\cite{Bienfait2020, kim2024}. Inspired by this prospect, recent experiments have realized qudits using photons with high orbital angular momentum \cite{Vaziri2003}, combining multiple photons~\cite{Lanyon2008,Luo2019} and through time-bin encoding \cite{NisbetJones2013}, as well as in other systems, such as superconducting circuits\cite{Qutrit2021, Goss2022, Superconducting2023PRL}, and trapped ions \cite{Qutrit_ion_2003,Practical_PRR_2020,ion_Ringbauer2022,Hrmo2023}.

Cold atoms in electronically high-lying Rydberg states (principal quantum number $n\gg 1$) have emerged as a new platform to realize quantum simulation and computation tasks~\cite{Saffman2010}, due to their strong and long-range two-body interactions, long lifetimes $\tau$ ($\tau\propto n^3$, typically $10\sim 100\,\mu$s for alkali atoms) and high tunability with external electromagnetic fields. Rapid developments have been made by using individual Rydberg atoms in carrying out quantum simulation and computation  tasks~\cite{bernienProbingManybodyDynamics2017,leseleucObservationSymmetryprotectedTopological2019,schollQuantumSimulation2D2021,ebadiQuantumPhasesMatter2021,bornet2023,bluvstein2024}. Complementary to approaches using individual atoms, a different route is to implement the collectively encoded qubits~\cite{brionQuantumComputingCollective2007}, where the strong-interaction induced excitation blockade enables collective encoding through storing information into superposed Rydberg states~\cite{Dipole_Lukin_2001,Dephasing_Bariani_2012, Dipolar_Tresp_2015,khazaliFastMultiqubitGates2020,wuSystematicErrorTolerantMultiqubitHolonomic2021,fan2023ManipulationSingleStoredphoton}.  ~\cite{dudin2012StronglyInteractingRydberg}. Moreover, large transition dipole moments allow fast Rabi flopping on nanosecond timescales by microwave (MW) fields, and coherent manipulation of Rydberg atoms~\cite{Coherent_Barredo_2015}.
While coherent rotation and read-out of collective qubits have been demonstrated~\cite{spong2021collectively}, implementation of higher dimensional computational units in a Rydberg atom ensemble, such as qutrits, is an unexplored research topic.

\begin{figure*}[thb]
	\centering
	\includegraphics[width=1\textwidth]{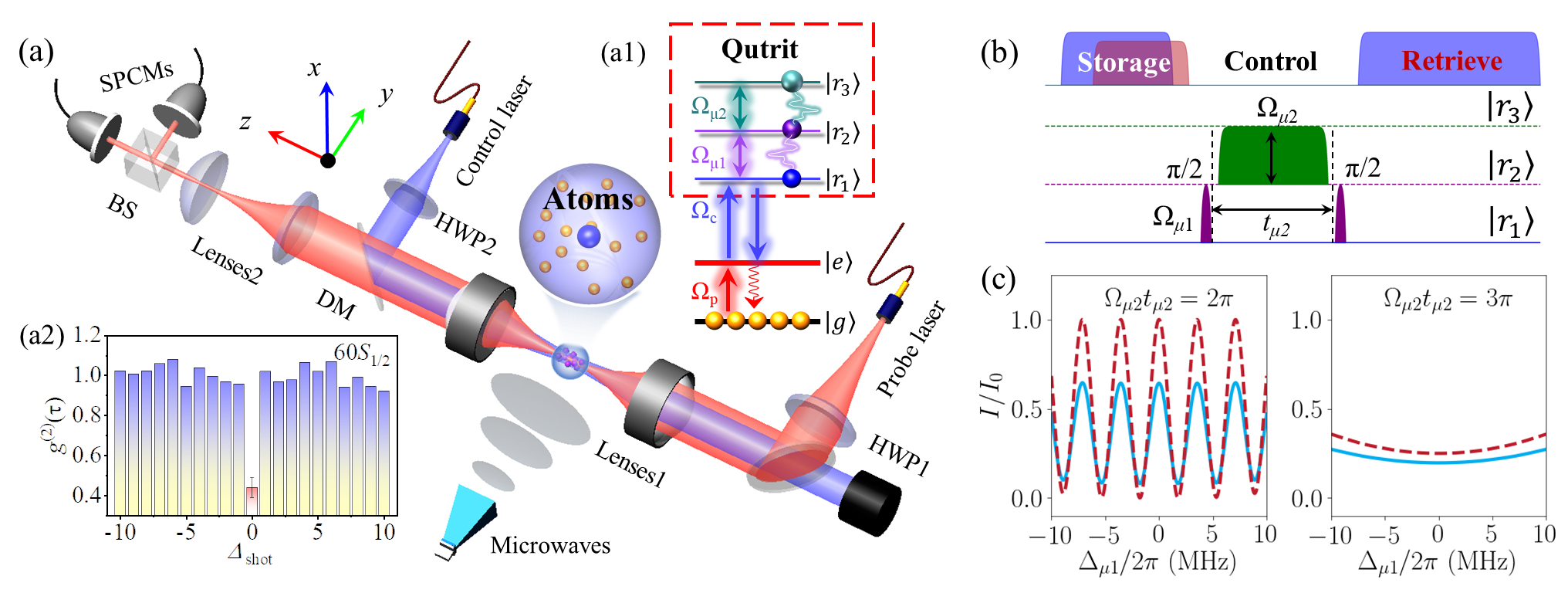} 
	\caption{\textbf{The setting}. (a) The experimental apparatus. A probe and control laser counter-propagate in an ultracold atom ensemble. Atoms are excited from ground state $\vert g\rangle$ to Rydberg state $\vert r_1\rangle$ via an intermediate state $\vert e\rangle$.  The relevant atomic levels are shown in (a1). By storing a probe photon in a Rydberg state $\vert r_1\rangle$ as Rydberg polaritons, we employ MW fields to control and manipulate collective Rydberg states that form state vectors of qutrit. An $\mu1$ MW field (Rabi frequency $\Omega_{\mu1}$) drives the transition $\vert {r_1}\rangle\to\vert {r_2}\rangle$. The $\mu2$ MW field (Rabi frequency $\Omega_{\mu2}$) is applied to connect to the third Rydberg state $\vert {r_3}\rangle$. (a2) The measured $g^{(2)}(\tau)$ of the retrieval signal in $\vert r_1\rangle$. (b) Time sequence of a Ramsey interferometer by modulating the $\mu2$-field. A visualisation of this sequence is shown in the supplementary material, based on the `octant plot' framework \cite{festenstein2023arbitrary}. (c) Interference fringes as a function of $\Delta_{\mu1}$ with $\Omega_{\mu2}t_{\mu2}=2\pi$ (left) and $3\pi$ (right), respectively. The red dashed line shows a calculation based on Eq.~(\ref{I_0}). The blue solid line represents results taking into account the dissipation. See text for details.}
	\label{fig:1}
\end{figure*}

In this work, we experimentally create and manipulate \textit{a single qutrit}
encoded in collective Rydberg excitation in an ensemble of cold $^{87}$Rb atoms (Fig.~\ref{fig:1}). A key to our collectively encoded qutrit is to prepare a Rydberg polariton in a superposition state of three different Rydberg levels $\vert r_\alpha\rangle$ ($\alpha = 1,2,3$), depicted in Fig.~\ref{fig:1}(a1). Using electromagnetically induced transparency (EIT)~\cite{pritchardCooperativeAtomLightInteraction2010,petrosyanElectromagneticallyInducedTransparency2011,gorshkovPhotonPhotonInteractionsRydberg2011}, a weak probe photon field is coherently transferred into a collective, singly excited Rydberg polariton state (see $g^{(2)}(0)$ shown in Fig.~\ref{fig:1}(a2)) exploiting the Rydberg blockade~\cite{maxwellStorageControlOptical2013}. We demonstrate coherent control among state vectors of the qutrit by MW fields, and determine occupations of the state vector through a time-bin encoded measurement. Subsequently, coherence of the qutrit is obtained through an MW-driven Ramsey protocol~\cite{Jiao2020,spong2021collectively}. We adopt an effective field model~\cite{otterbachWignerCrystallizationSingle2013,gullansEffectiveFieldTheory2016} to determine the dynamics of the qutrit, and derive analytical visibility of Ramsey fringes. We show that the qutrit operation by MW fields is coherent, leading to high visibility that agrees with the theoretical prediction. The joint experimental and theoretical investigation opens a way to exploiting the potential of collectively Rydberg qutrit in quantum information processing~\cite{brionQuantumComputingCollective2007}, and compliments to the atom array approach~\cite{bernienProbingManybodyDynamics2017,browaeysManybodyPhysicsIndividually2020}. 

\textit {Model---}
We consider a single qutrit whose state vectors are encoded in collectively excited Rydberg states $\vert R_\alpha\rangle=1/\sqrt{N}\sum_{j=1}^N 
{\rm  e}^{{\rm i}
	\boldsymbol{k}\cdot \boldsymbol{r}_j}
\vert  g\rangle_1\ldots \vert r_\alpha\rangle_j \ldots \vert  g\rangle_N$, where $\boldsymbol{k}$ and $N$ are the effective wave vector and number of atoms of the collective (spinwave) state. Here $\boldsymbol{k}\cdot\boldsymbol{r}_j$ gives the phase for an atom at position $\boldsymbol{r}_j$. Two microwave fields, $\mu1$ and $\mu2$, couple collective states $\lvert R_1\rangle \leftrightarrow  \lvert R_2\rangle$ and $\lvert R_2\rangle \leftrightarrow  \lvert R_3\rangle$ coherently. $\Delta_{\mu\alpha}(t)$ and $\Omega_{\mu\alpha}(t)$
are the detuning and Rabi frequency, respectively. Coupling of the weakly excited collective state is described by  Hamiltonian~\cite{Fleischhauer2000}, 
\begin{eqnarray} \label{Hamiltonian}
		&\hat{H}&=\int d{\mathbf r}\left[
		-\Delta_{\mu1}(t) \hat{\Psi}_2^\dagger({\mathbf r})\hat{\Psi}_2({\mathbf r})-\Delta_{\mu2}(t) \hat{\Psi}_3^\dagger({\mathbf r})\hat{\Psi}_3({\mathbf r})\right.\nonumber\\
		&+&\left.\frac{\Omega_{\mu1}(t)}{2}\hat{\Psi}_2^\dagger({\mathbf r})\hat{\Psi}_1({\mathbf r})+\frac{\Omega_{\mu2}(t)}{2}\hat{\Psi}_3^\dagger({\mathbf r})\hat{\Psi}_2({\mathbf r})+{\rm H.c.}\right],\nonumber
\end{eqnarray}
where we have used continuous field operators $\hat{\Psi}_\alpha({\mathbf r})=\vert G\rangle\langle R_{\alpha}\vert$ with $\vert G\rangle = \vert  g\rangle_1\ldots  \ldots \vert  g\rangle_N$ to be the $N$ atom ground state. Operator $\hat{\Psi}_\alpha({\mathbf r})$ annihilates an excitation in state $|r_\alpha\rangle$ at location $\mathbf{r}$. 
These operators satisfy the equal time, bosonic commutation relation, $[\hat{\Psi}_\alpha({\mathbf r}), \hat{\Psi}_\beta^\dagger({\mathbf r}^\prime)]=\delta_{\alpha\beta}\delta({\mathbf r}-{\mathbf r}^\prime)$.  The effective field theoretical description is a good approximation in high density gases~\cite{Fleischhauer2000}. 

Coherence of a qutrit can be characterized by a Ramsey interferometer which constitutes two $\pi$/2 MW ($\mu 1$) pulses 
 (see Fig.~\ref{fig:1}(b) for the Ramsey sequence).
%
Phase differences of the three states are modulated by varying duration $t_{\mu2}$ of MW field $\mu2$.
The Ramsey process is described by rotation matrices derived analytically (See Supplemental Material ({\bf SM}) for details). 
Subsequently, the population in $\vert R_1\rangle$ is read out by coupling the qutrit back to the excited state with the control laser~\cite{spong2021collectively}. The resulting photon count $I$ is analytically evaluated, 
\begin{eqnarray}\label{I_0}
	I
	& = & I_0\left(|A|^2+|B|^2\cos^2\frac{\Omega_{\mu2}t_{\mu2}}{2}+C\cos\frac{\Omega_{\mu2}t_{\mu2}}{2}\right)
\end{eqnarray}
where $A=e^{i {\Delta_{\mu1}}t_{\mu1}}(\cos \frac{1}{2} \Phi-i \Delta_{\mu1} t_{\mu1} \sin \frac{1}{2} \Phi/\Phi)^2$, $B=-e^{i \Delta_{\mu1}t_{\rm total} }\pi ^2 \sin ^2\left(\frac{1}{2} \Phi\right)/(4 \Delta_{\mu1}^2 t_{\mu1}^2+\pi ^2)$, and $C=A B^\ast+A^\ast B$ represent the interference terms with $\Phi=(\Delta_{\mu1}^2 t_{\mu1}^2+\pi ^2/4)^{1/2}$ and   
the total time $t_{\rm total}\simeq t_{\mu1}+ t_{\mu2}$.
In the absence of the $\mu2$-field, $\Omega_{\mu2} = 0$, Eq.~(\ref{I_0}) reduces to the conventional Ramsey fringe, $	I
=  I_0|A+B|^2$, in the $\{|R_1\rangle, |R_2\rangle\}$ basis. Turning on the $\mu2$-field, the Ramsey interferometer displays perfect fringes for $\Omega_{\mu2}t_{\mu2}=2k\pi$, and no fringes when  $\Omega_{\mu2}t_{\mu2}=(2k+1)\pi$ ($k = 0, 1, 2 ...$). As an example, fringes as a function of  detuning $\Delta_{\mu1}$ for 
$\Omega_{\mu2}t_{\mu2}=2\pi$ and $3\pi$ are shown as the red dashed line in Fig.~\ref{fig:1}(c), respectively.

 The qutrit is robust against typical dissipation (e.g. atomic decay, dephasing processes, or atomic collision) due to its collective nature. Taking these effects into account, we investigate the robustness of the qutrit based on a quantum master equation, $\dot{\hat{\rho}}=-i[H,\hat{\rho}]+{\mathcal L}_{\rm Decay}(\hat{\rho})+{\mathcal L}_{\rm Deph}(\hat{\rho})$, where Lindblad operators ${\mathcal L}_{\rm Decay}(\hat{\rho})$ and ${\mathcal L}_{\rm Deph}(\hat{\rho})$ describe decay and dephasing processes (See \textbf{SM} for details). 
Compared to the dissipation free case Eq.~(\ref{I_0}), the interference fringe is preserved, though its amplitude becomes lower due to the various dissipative processes, shown as the blue solid line in Fig.~\ref{fig:1}(c). This is in contrast to that of single-atom
qutrits, where all information is lost due to, e.g., loss of the atom.

 \begin{figure}[thb]
    \centering	\includegraphics[width=1
\linewidth]{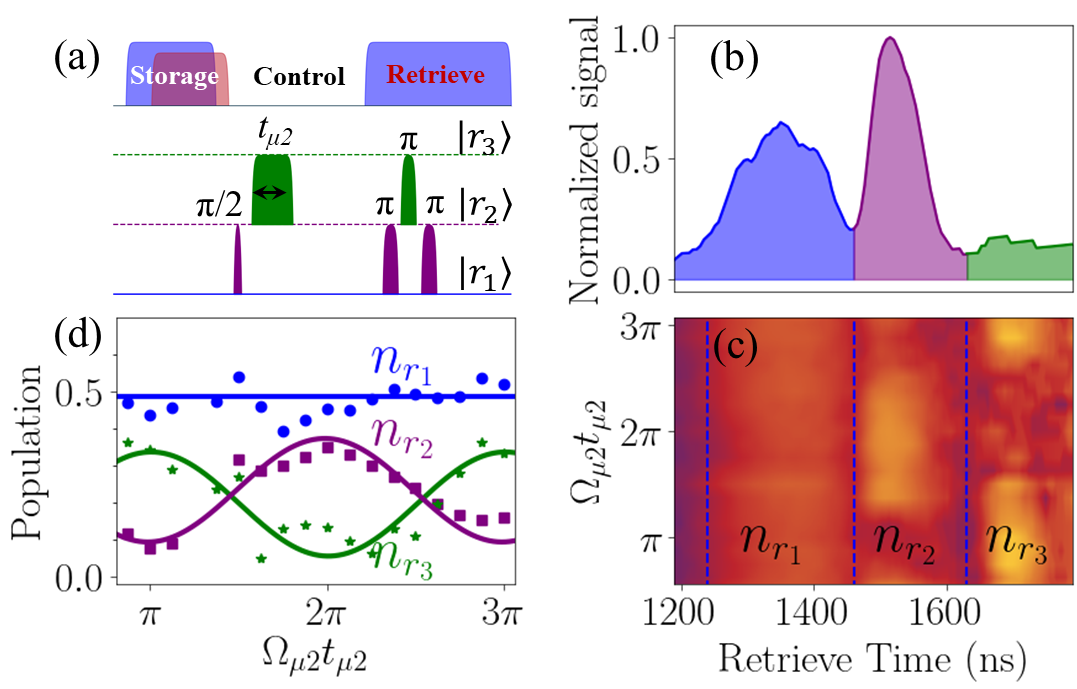}
    \caption{\textbf{Read-out of the qutrit vectors.} (a) Timing sequences for performing the control and read-out of qutrit. (b) The read-out of population in states $\vert {R_1}\rangle$ (blue area), $\vert {R_2}\rangle$ (purple area) and $\vert {R_3}\rangle$ (green area) are about 47.1~\%, 38.6~\% and 14.2~\%, respectively.
 In this experiment, we set $\Omega_{\mu2}t_{\mu2}=2\pi$.
	(c) The time evolution of the populations in three different Rydberg states for $\Omega_{\mu2} = 2\pi\times12.5$~MHz. Changing time $t_{\mu2}$, populations in $\vert R_2\rangle$ and $\vert R_3\rangle$ oscillate. 
	(d) Population of $\vert {R_\alpha}\rangle$ integrated over the retrieve time. It is shown the Rabi oscillation between states $\vert {R_2}\rangle$ and $\vert {R_3}\rangle$. The solid lines are sinusoidal fittings for guiding eyes.}
\label{fig:3}
\end{figure}

\textit {Experimental preparation, control and read-out of a qutrit---}
In our experiment, up to 1000 $^{87}$Rb atoms are trapped by an 862~nm optical tweezer with a $1/e^2$ waist of $w_0=4.5~{\rm \mu m}$ and the trap depth of $\sim 1.0~{\rm mK}$ (see Fig.~\ref{fig:1}(a)). The atoms are initially pumped into the ground state $\vert g\rangle=\vert{5S_{1/2}, F = 2, m_F = 2}\rangle$. 
A 780~nm probe laser couples state $|g\rangle$ to an excited state $\vert e\rangle = \vert{5P_{3/2}, F' = 3, m'_F = 3}\rangle$, and a strong 480 nm control beam is resonant coupling the transition of $|e\rangle$ $\to $ $\vert r_1\rangle = \vert{60S_{1/2}}\rangle$.
The probe and control lasers are opposite circularly polarized and counter-propagating through the atomic ensemble (size of 1.5~${\rm \mu m}$ in radial and 25~${\rm \mu m}$ in axial)  forming an ellipsoidal excitation region.
By reducing the intensity of the control field over a time of 700 ns before the probe is switched off, a probe photon is stored as a collective polariton state
$\vert R_1\rangle$, where a single Rydberg excitation in $\vert r_1\rangle$ is shared by $N$ atoms~\cite{Storage_Maxwell_2013}. The phase of the collective state is important, which ensures the read-out photons in a well-defined spatial mode as the original input~\cite{Dudin2012,Ornelas-Huerta2020}. 

Once state $\vert R_1\rangle$ is prepared, it can be read out when the control field is applied to drive the transition $\vert {r_1}\rangle\to\vert {e}\rangle$.
State $\vert {e}\rangle$ decays spontaneously to the ground state $\vert {g}\rangle$ on a time scale of $10~{\rm ns}$ in which a single photon is emitted in the same mode as the input probe photon. This allows us to measure the second-order correlation of the retrieved photons using a standard Hanbury-Brown-Twiss (HBT) technique. Fig.~\ref{fig:1}(a2) shows the correlation $g^{(2)}(0) = 0.45\pm0.05$, indicating that we prepare, largely, a single excitation in state $\vert R_1\rangle$.  
The blockade radius (about 8$\mu$m) for \textit{n} = 60 is smaller than the 
axis size of the 
atomic ensemble. This allows two or more Rydberg excitation, affecting the correlation $g^{(2)}(0)$. The multiple Rydberg excitations could also affect the operations of the qutrit (see discussion below). On the other hand, the purity of the qutrit can be further enhanced, e.g. by filtering the background, reducing the size of the atomic ensemble or adopting higher-lying Rydberg states~\cite{Jiao2020,Ornelas-Huerta2020}. 

The collectively encoded qutrit can be controlled coherently with MW fields. Starting with state $|R_1\rangle$, the first $\pi$/2 pulse of the ${\mu1}$-field is calibrated to split it into a superposition of states $\vert {R_1}\rangle$ and a new collective state $\vert {R_2}\rangle$ (composed by a single Rydberg excitation in state $\vert r_2\rangle = \vert{59{P}_{3/2}}\rangle$), with resonant frequency 18.51~GHz. We then apply the ${\mu2}$ field to continuously drive  transition $\vert {R_2}\rangle\rightarrow \vert R_3\rangle$ for  duration $t_{\mu 2}$, where $\vert R_3\rangle$ is 
 a collective excitation in Rydberg state $\vert{59{S}_{1/2}}\rangle$. The resonant frequency for this case is 18.23~GHz. The sequence of the collective state control is shown in Fig.~\ref{fig:3}(a).
 
By applying the MW fields, the qutrit is a superposition of all three state vectors $\vert R_{\alpha}\rangle$. We can determine the occupation by retrieving the collective states consisting of the qutrit individually. State $\vert R_1\rangle$ is retrieved directly by turning on the control fields, as discussed previously. We can then sequentially read states $\vert R_2\rangle$ and $\vert R_3\rangle$. First, we apply a $\pi$ pulse of the $\mu1$-field to pump populations in state $\vert {R_2}\rangle$ back to state $\vert {R_1}\rangle$, which is immediately read out. Second, two $\pi$ pulses of the $\mu1$- and  $\mu2$-fields are sequentially applied to couple transition $\vert {R_3}\rangle\to\vert {R_2}\rangle\to\vert {R_1}\rangle$, such that we read out the population in state $\vert {R_3}\rangle$. In Fig.~\ref{fig:3}(b), 
we show populations in states $\vert {R_1}\rangle$ (blue area), $\vert {R_2}\rangle$ (purple area) and $\vert {R_3}\rangle$ (green area) for $\Omega_{\mu2}t_{\mu2}=2\pi$. The data is the statistics of one million sequences. 
The retrieved populations of the three states are 47.1~\%, 38.6~\% and 14.2~\%, diverging from the ideal case (50\%, 50\%, and 0\%) due to the dephasing caused by the time delay in the repumping process. The total sequence is less than 1.8~$\mu s$.

We control the occupation of the qutrit state by the MW fields. For example, Rabi oscillations between states $\vert {R_2}\rangle\rightarrow\vert {R_3}\rangle$ are induced by increasing $t_{\mu2}$ with fixed  $\Omega_{\mu2} = 2\pi\times12.5$\,MHz. 
As illustrated in Fig.~\ref{fig:3}(c), retrieved photon signals evolve with $\Omega_{\mu2}t_{\mu2}$.
After integrating over the retrieve time, the time evolution of the population in three Rydberg states is obtained (see Fig.~\ref{fig:3}(d)). 
As expected, the population in state $\vert {R_1}\rangle$ remains around $50\%$, while coherent Rabi oscillations are driven between the states $\vert {R_2}\rangle$ and $\vert {R_3}\rangle$.
The oscillation is maintained for 1.5 periods before noticeable decay. This decay occurs because we are continuously reading out the population, which is dissipative.

\begin{figure}[thb]
	\centering
	\includegraphics[width=1\linewidth]{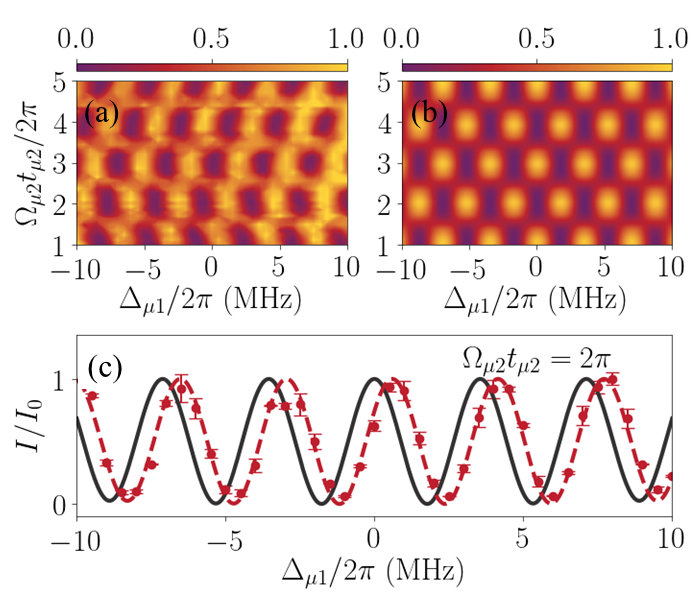}
	\caption{\textbf{Illustrating coherence of the prepared qutrit.} (a) Measured Ramsey fringes as a function of both $\Delta_{\mu1}$ and 
 $\Omega_{\mu2}$ with $t_{\mu 2}$ = 250 ns. Each data is normalized to its maximum. (b) Theoretical Ramsey fringes calculated with three-level rotation matrices based on Eq.~(\ref{I_0}). (c) The measured interference fringes (red dots) as a function of $\Delta_{\mu1}$ with $\Omega_{\mu2}t_{\mu2}=2\pi$. The black solid line shows the theoretical calculation. The red dashed line represents the results with Rydberg interactions.}
	\label{fig:5}
\end{figure}

\textit {Coherence of the Rydberg qutrit---} 
With the developed tools to prepare and manipulate the qutrit, we now investigate its coherence, and compare with the theoretical analysis. In a Ramsey sequence (see timing depicted in Fig.~\ref{fig:1}(b)), we first  measure interference fringes by varying $\Delta_{\mu1}$ and $\Omega_{\mu2}t_{\mu2}$. 
Here, we fix the pulse duration of the  $\mu2$-field  $t_{\mu2} = 250~{\rm ns} $ and increase Rabi frequency $\Omega_{\mu2}$. Experimental results are the statistics of one million sequences and are shown in Fig.~\ref{fig:5}(a). We obtain clear interference fringes by varying $\Delta_{\mu1}$ (or $\Omega_{\mu2}$), illustrating coherence of the prepared qutrit.

Compared to the theoretical prediction based on Eq.~(\ref{I_0})  (shown in Fig.~\ref{fig:5}(b)), one finds that the experimental and theoretical results agree well overall. A careful inspection shows that there is a phase deviation between the theoretical prediction and experimental data. This can be seen when comparing data where fringes are apparent, for example, at $\Omega_{\mu2} t_{\mu2}=K\pi$ with $K$ to be positive integers. One example with $\Omega_{\mu2}t_{\mu2}=2\pi$ is displayed in Fig.~\ref{fig:5}(c). We obtain a phase difference, about 3~MHz, between the experiment data and simulation (black solid line). This phase could be attributed to multiple Rydberg excitations during the preparation stage (see Fig.~\ref{fig:1}(a2)). Once two or more Rydberg excitations are excited, van der Waals interactions are expected among the collective Rydberg states. This leads to level shifts and hence a non-negligible phase difference with respect to the interaction free case (see Eq.~(\ref{I_0})). To verify this, we numerically simulate the master equation (see \textbf{SM} for simulation details) taking the double excitation into account. We can see that the interaction does not modify oscillation periods or amplitudes significantly. An excellent agreement between the theory (red dashed line) and experiment (dots) confirms that the Rydberg interaction plays an important role here. The Rydberg interaction would permit us to realize quantum logical operations, such as entangled gates with two collective ensembles~\cite{brionQuantumComputingCollective2007}. To suppress the interaction effect, one could consider even weaker probe fields, decrease spatial sizes of the ensemble to be smaller than the blockade sphere, or excite Rydberg states with larger $n$. Thus the purity of the qutrit can be further enhanced.  One possible drawback in higher $n$ Rydberg states is that they are much more sensitive to residual electromagnetic fields and blackbody radiation~\cite{beterov2009, zeiher2016, deleseleuc2018}, which affects the fidelity of rotation operations and visibility of Ramsey fringes.

\begin{figure}[thb]
    \centering	\includegraphics[width=1\linewidth]{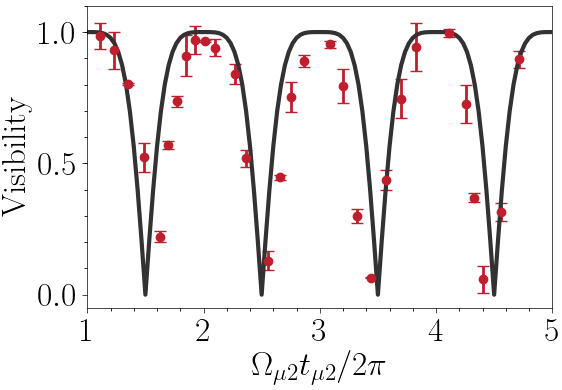}
	\caption{\textbf{Restoring quantum coherence of the qutrit.} Data points show the visibility of the Ramsey fringes as a function of the $\Omega_{\mu2}$. The solid line is the theoretical result determined by Eq.~(\ref{visibility}). An excellent agreement between the experiment and theory can be seen. }
	\label{fig:4}
\end{figure}

The $\mu2$-field modifies the Ramsey fringe, whose visibility is in fact evaluated analytically (See \textbf{SM} for details),
\begin{eqnarray}\label{visibility}
	V
	& = & \left\vert
	\frac{2\cos(\Omega_{\mu2}t_{\mu2}/2)}{ 1+\cos^2(\Omega_{\mu2}t_{\mu2}/2)}\right\vert,
\end{eqnarray}
which shows that the visibility oscillates between $0$ and $1$. This takes place as $\Omega_{\mu 2}t_{\mu2}$ changes in one period $2\pi$. 
By fitting experimental data, we extract the visibility of the Ramsey interferometer as a function of  $\Omega_{\mu2}t_{\mu2}$ (red dots). The experimental results are consistent with Eq.~(\ref{visibility}) (solid black line).
The visibility goes zero when the population in ${R_3}$ is a maximum, i.e., the rotation in $\{|R_2\rangle, |R_3\rangle\}$ space is an odd multiple of $\pi$. Compared with direct read-out of the information in state $\vert R_2\rangle$, the Ramsey interferometer provides a non–destructive detection of the Rydberg state, which is preferable for quantum information processing. 

\textit {Conclusion and outlook---}
We have experimentally realized the storage, MW control and fast read-out of a single photonic qutrit in an ultracold atomic ensemble for the first time, and developed an effective field model to describe qutrit dynamics.
A Ramsey interferometer is used to measure the coherence of the qutrit, which demonstrates the coherent connectivity among the high qutrit space. The Ramsey fringe visibility oscillates with the Rabi frequency of the MW field in excellent agreement with our theoretical simulation. Our experimental and theoretical study demonstrate the high fidelity control and robustness of the collectively encoded Rydberg qutrit. The qutrit can be used to probe novel topological physics, such as high Chern numbers~\cite{tanTopologicalMaxwellMetal2018}. The collectively encoded qutrit is robust against particle loss~\cite{spong2021collectively}, which opens a route to explore topological effects between arrays of qutrits stored in multiple clouds~\cite{paredes-barato2014AllOpticalQuantumInformationb, Busche2017, khazali2019PolaritonExchangeInteractions}. Using the higher Hibert dimension of the qutrit will further allow efficient quantum algorithms \cite{Fedorov2012,Gokhale2019}, increased security \cite{Vaziri2003,Lanyon2008,Luo2019}, novel approaches to testing fundamental quantum mechanics~\cite{collins2002BellInequalitiesArbitrarily}, and quantum sensing applications~\cite{yang2024a}.

\textit {Acknowledgment---} C.S.A. acknowledges financial support from the UKRI, EPSRC grant reference number EP/V030280/1 (“Quantum optics using Rydberg polaritons”). Y.C.J., Z.Y.B. and J.M.Z. acknowledge financial support from the National Natural Science Foundation of China (No. 12241408, 12120101004, U2341211, 62175136 and 12274131). W.L. acknowledges financial support from the EPSRC  (Grant No. EP/W015641/1) and the Going Global Partnerships Programme of the British Council (Contract No. IND/CONT/G/22-23/26).

\bibliography{main}

\begin{thebibliography}{60}%
\makeatletter
\providecommand \@ifxundefined [1]{%
 \@ifx{#1\undefined}
}%
\providecommand \@ifnum [1]{%
 \ifnum #1\expandafter \@firstoftwo
 \else \expandafter \@secondoftwo
 \fi
}%
\providecommand \@ifx [1]{%
 \ifx #1\expandafter \@firstoftwo
 \else \expandafter \@secondoftwo
 \fi
}%
\providecommand \natexlab [1]{#1}%
\providecommand \enquote  [1]{``#1''}%
\providecommand \bibnamefont  [1]{#1}%
\providecommand \bibfnamefont [1]{#1}%
\providecommand \citenamefont [1]{#1}%
\providecommand \href@noop [0]{\@secondoftwo}%
\providecommand \href [0]{\begingroup \@sanitize@url \@href}%
\providecommand \@href[1]{\@@startlink{#1}\@@href}%
\providecommand \@@href[1]{\endgroup#1\@@endlink}%
\providecommand \@sanitize@url [0]{\catcode `\\12\catcode `\$12\catcode `\&12\catcode `\#12\catcode `\^12\catcode `\_12\catcode `\%12\relax}%
\providecommand \@@startlink[1]{}%
\providecommand \@@endlink[0]{}%
\providecommand \url  [0]{\begingroup\@sanitize@url \@url }%
\providecommand \@url [1]{\endgroup\@href {#1}{\urlprefix }}%
\providecommand \urlprefix  [0]{URL }%
\providecommand \Eprint [0]{\href }%
\providecommand \doibase [0]{https://doi.org/}%
\providecommand \selectlanguage [0]{\@gobble}%
\providecommand \bibinfo  [0]{\@secondoftwo}%
\providecommand \bibfield  [0]{\@secondoftwo}%
\providecommand \translation [1]{[#1]}%
\providecommand \BibitemOpen [0]{}%
\providecommand \bibitemStop [0]{}%
\providecommand \bibitemNoStop [0]{.\EOS\space}%
\providecommand \EOS [0]{\spacefactor3000\relax}%
\providecommand \BibitemShut  [1]{\csname bibitem#1\endcsname}%
\let\auto@bib@innerbib\@empty
\bibitem [{\citenamefont {Blok}\ \emph {et~al.}(2021{\natexlab{a}})\citenamefont {Blok}, \citenamefont {Ramasesh}, \citenamefont {Schuster}, \citenamefont {O'Brien}, \citenamefont {Kreikebaum}, \citenamefont {Dahlen}, \citenamefont {Morvan}, \citenamefont {Yoshida}, \citenamefont {Yao},\ and\ \citenamefont {Siddiqi}}]{blok2021}%
  \BibitemOpen
  \bibfield  {author} {\bibinfo {author} {\bibfnamefont {M.~S.}\ \bibnamefont {Blok}}, \bibinfo {author} {\bibfnamefont {V.~V.}\ \bibnamefont {Ramasesh}}, \bibinfo {author} {\bibfnamefont {T.}~\bibnamefont {Schuster}}, \bibinfo {author} {\bibfnamefont {K.}~\bibnamefont {O'Brien}}, \bibinfo {author} {\bibfnamefont {J.~M.}\ \bibnamefont {Kreikebaum}}, \bibinfo {author} {\bibfnamefont {D.}~\bibnamefont {Dahlen}}, \bibinfo {author} {\bibfnamefont {A.}~\bibnamefont {Morvan}}, \bibinfo {author} {\bibfnamefont {B.}~\bibnamefont {Yoshida}}, \bibinfo {author} {\bibfnamefont {N.~Y.}\ \bibnamefont {Yao}},\ and\ \bibinfo {author} {\bibfnamefont {I.}~\bibnamefont {Siddiqi}},\ }\bibfield  {title} {\bibinfo {title} {Quantum {{Information Scrambling}} on a {{Superconducting Qutrit Processor}}},\ }\href {https://doi.org/10.1103/PhysRevX.11.021010} {\bibfield  {journal} {\bibinfo  {journal} {Phys. Rev. X}\ }\textbf {\bibinfo {volume} {11}},\ \bibinfo {pages} {021010} (\bibinfo {year} {2021}{\natexlab{a}})}\BibitemShut
  {NoStop}%
\bibitem [{\citenamefont {Mackeprang}\ \emph {et~al.}(2023)\citenamefont {Mackeprang}, \citenamefont {Bhatti}, \citenamefont {Hoban},\ and\ \citenamefont {Barz}}]{mackeprang2023}%
  \BibitemOpen
  \bibfield  {author} {\bibinfo {author} {\bibfnamefont {J.}~\bibnamefont {Mackeprang}}, \bibinfo {author} {\bibfnamefont {D.}~\bibnamefont {Bhatti}}, \bibinfo {author} {\bibfnamefont {M.~J.}\ \bibnamefont {Hoban}},\ and\ \bibinfo {author} {\bibfnamefont {S.}~\bibnamefont {Barz}},\ }\bibfield  {title} {\bibinfo {title} {The power of qutrits for non-adaptive measurement-based quantum computing},\ }\href {https://doi.org/10.1088/1367-2630/acdf77} {\bibfield  {journal} {\bibinfo  {journal} {New J. Phys.}\ }\textbf {\bibinfo {volume} {25}},\ \bibinfo {pages} {073007} (\bibinfo {year} {2023})}\BibitemShut {NoStop}%
\bibitem [{\citenamefont {Wang}\ \emph {et~al.}(2020)\citenamefont {Wang}, \citenamefont {Hu}, \citenamefont {Sanders},\ and\ \citenamefont {Kais}}]{wang2020Qudits}%
  \BibitemOpen
  \bibfield  {author} {\bibinfo {author} {\bibfnamefont {Y.}~\bibnamefont {Wang}}, \bibinfo {author} {\bibfnamefont {Z.}~\bibnamefont {Hu}}, \bibinfo {author} {\bibfnamefont {B.~C.}\ \bibnamefont {Sanders}},\ and\ \bibinfo {author} {\bibfnamefont {S.}~\bibnamefont {Kais}},\ }\bibfield  {title} {\bibinfo {title} {Qudits and {{High-Dimensional Quantum Computing}}},\ }\href {https://doi.org/10.3389/fphy.2020.589504} {\bibfield  {journal} {\bibinfo  {journal} {Front. Phys.}\ }\textbf {\bibinfo {volume} {8}},\ \bibinfo {pages} {589504} (\bibinfo {year} {2020})}\BibitemShut {NoStop}%
\bibitem [{\citenamefont {Rambach}\ \emph {et~al.}(2021)\citenamefont {Rambach}, \citenamefont {Qaryan}, \citenamefont {Kewming}, \citenamefont {Ferrie}, \citenamefont {White},\ and\ \citenamefont {Romero}}]{Robust_2021_PRL}%
  \BibitemOpen
  \bibfield  {author} {\bibinfo {author} {\bibfnamefont {M.}~\bibnamefont {Rambach}}, \bibinfo {author} {\bibfnamefont {M.}~\bibnamefont {Qaryan}}, \bibinfo {author} {\bibfnamefont {M.}~\bibnamefont {Kewming}}, \bibinfo {author} {\bibfnamefont {C.}~\bibnamefont {Ferrie}}, \bibinfo {author} {\bibfnamefont {A.~G.}\ \bibnamefont {White}},\ and\ \bibinfo {author} {\bibfnamefont {J.}~\bibnamefont {Romero}},\ }\bibfield  {title} {\bibinfo {title} {Robust and efficient high-dimensional quantum state tomography},\ }\href {https://doi.org/10.1103/PhysRevLett.126.100402} {\bibfield  {journal} {\bibinfo  {journal} {Phys. Rev. Lett.}\ }\textbf {\bibinfo {volume} {126}},\ \bibinfo {pages} {100402} (\bibinfo {year} {2021})}\BibitemShut {NoStop}%
\bibitem [{\citenamefont {Fedorov}\ \emph {et~al.}(2012)\citenamefont {Fedorov}, \citenamefont {Steffen}, \citenamefont {Baur}, \citenamefont {da~Silva},\ and\ \citenamefont {Wallraff}}]{Fedorov2012}%
  \BibitemOpen
  \bibfield  {author} {\bibinfo {author} {\bibfnamefont {A.}~\bibnamefont {Fedorov}}, \bibinfo {author} {\bibfnamefont {L.}~\bibnamefont {Steffen}}, \bibinfo {author} {\bibfnamefont {M.}~\bibnamefont {Baur}}, \bibinfo {author} {\bibfnamefont {M.~P.}\ \bibnamefont {da~Silva}},\ and\ \bibinfo {author} {\bibfnamefont {A.}~\bibnamefont {Wallraff}},\ }\bibfield  {title} {\bibinfo {title} {Implementation of a toffoli gate with superconducting circuits},\ }\href {https://doi.org/10.1038/nature10713} {\bibfield  {journal} {\bibinfo  {journal} {Nature}\ }\textbf {\bibinfo {volume} {481}},\ \bibinfo {pages} {170} (\bibinfo {year} {2012})}\BibitemShut {NoStop}%
\bibitem [{\citenamefont {Gokhale}\ \emph {et~al.}(2019)\citenamefont {Gokhale}, \citenamefont {Baker}, \citenamefont {Duckering}, \citenamefont {Brown}, \citenamefont {Brown},\ and\ \citenamefont {Chong}}]{Gokhale2019}%
  \BibitemOpen
  \bibfield  {author} {\bibinfo {author} {\bibfnamefont {P.}~\bibnamefont {Gokhale}}, \bibinfo {author} {\bibfnamefont {J.~M.}\ \bibnamefont {Baker}}, \bibinfo {author} {\bibfnamefont {C.}~\bibnamefont {Duckering}}, \bibinfo {author} {\bibfnamefont {N.~C.}\ \bibnamefont {Brown}}, \bibinfo {author} {\bibfnamefont {K.~R.}\ \bibnamefont {Brown}},\ and\ \bibinfo {author} {\bibfnamefont {F.~T.}\ \bibnamefont {Chong}},\ }\bibfield  {title} {\bibinfo {title} {Asymptotic improvements to quantum circuits via qutrits},\ }in\ \href {https://doi.org/10.1145/3307650.3322253} {\emph {\bibinfo {booktitle} {Proceedings of the 46th International Symposium on Computer Architecture}}},\ \bibinfo {series and number} {ISCA ’19}\ (\bibinfo  {publisher} {Association for Computing Machinery},\ \bibinfo {address} {New York, NY, USA},\ \bibinfo {year} {2019})\ pp.\ \bibinfo {pages} {554--566}\BibitemShut {NoStop}%
\bibitem [{\citenamefont {Vaziri}\ \emph {et~al.}(2003)\citenamefont {Vaziri}, \citenamefont {Pan}, \citenamefont {Jennewein}, \citenamefont {Weihs},\ and\ \citenamefont {Zeilinger}}]{Vaziri2003}%
  \BibitemOpen
  \bibfield  {author} {\bibinfo {author} {\bibfnamefont {A.}~\bibnamefont {Vaziri}}, \bibinfo {author} {\bibfnamefont {J.-W.}\ \bibnamefont {Pan}}, \bibinfo {author} {\bibfnamefont {T.}~\bibnamefont {Jennewein}}, \bibinfo {author} {\bibfnamefont {G.}~\bibnamefont {Weihs}},\ and\ \bibinfo {author} {\bibfnamefont {A.}~\bibnamefont {Zeilinger}},\ }\bibfield  {title} {\bibinfo {title} {Concentration of higher dimensional entanglement: Qutrits of photon orbital angular momentum},\ }\href {https://doi.org/10.1103/PhysRevLett.91.227902} {\bibfield  {journal} {\bibinfo  {journal} {Phys. Rev. Lett.}\ }\textbf {\bibinfo {volume} {91}},\ \bibinfo {pages} {227902} (\bibinfo {year} {2003})}\BibitemShut {NoStop}%
\bibitem [{\citenamefont {Lanyon}\ \emph {et~al.}(2008)\citenamefont {Lanyon}, \citenamefont {Weinhold}, \citenamefont {Langford}, \citenamefont {O'Brien}, \citenamefont {Resch}, \citenamefont {Gilchrist},\ and\ \citenamefont {White}}]{Lanyon2008}%
  \BibitemOpen
  \bibfield  {author} {\bibinfo {author} {\bibfnamefont {B.~P.}\ \bibnamefont {Lanyon}}, \bibinfo {author} {\bibfnamefont {T.~J.}\ \bibnamefont {Weinhold}}, \bibinfo {author} {\bibfnamefont {N.~K.}\ \bibnamefont {Langford}}, \bibinfo {author} {\bibfnamefont {J.~L.}\ \bibnamefont {O'Brien}}, \bibinfo {author} {\bibfnamefont {K.~J.}\ \bibnamefont {Resch}}, \bibinfo {author} {\bibfnamefont {A.}~\bibnamefont {Gilchrist}},\ and\ \bibinfo {author} {\bibfnamefont {A.~G.}\ \bibnamefont {White}},\ }\bibfield  {title} {\bibinfo {title} {Manipulating biphotonic qutrits},\ }\href {https://doi.org/10.1103/PhysRevLett.100.060504} {\bibfield  {journal} {\bibinfo  {journal} {Phys. Rev. Lett.}\ }\textbf {\bibinfo {volume} {100}},\ \bibinfo {pages} {060504} (\bibinfo {year} {2008})}\BibitemShut {NoStop}%
\bibitem [{\citenamefont {Luo}\ \emph {et~al.}(2019)\citenamefont {Luo}, \citenamefont {Zhong}, \citenamefont {Erhard}, \citenamefont {Wang}, \citenamefont {Peng}, \citenamefont {Krenn}, \citenamefont {Jiang}, \citenamefont {Li}, \citenamefont {Liu}, \citenamefont {Lu}, \citenamefont {Zeilinger},\ and\ \citenamefont {Pan}}]{Luo2019}%
  \BibitemOpen
  \bibfield  {author} {\bibinfo {author} {\bibfnamefont {Y.-H.}\ \bibnamefont {Luo}}, \bibinfo {author} {\bibfnamefont {H.-S.}\ \bibnamefont {Zhong}}, \bibinfo {author} {\bibfnamefont {M.}~\bibnamefont {Erhard}}, \bibinfo {author} {\bibfnamefont {X.-L.}\ \bibnamefont {Wang}}, \bibinfo {author} {\bibfnamefont {L.-C.}\ \bibnamefont {Peng}}, \bibinfo {author} {\bibfnamefont {M.}~\bibnamefont {Krenn}}, \bibinfo {author} {\bibfnamefont {X.}~\bibnamefont {Jiang}}, \bibinfo {author} {\bibfnamefont {L.}~\bibnamefont {Li}}, \bibinfo {author} {\bibfnamefont {N.-L.}\ \bibnamefont {Liu}}, \bibinfo {author} {\bibfnamefont {C.-Y.}\ \bibnamefont {Lu}}, \bibinfo {author} {\bibfnamefont {A.}~\bibnamefont {Zeilinger}},\ and\ \bibinfo {author} {\bibfnamefont {J.-W.}\ \bibnamefont {Pan}},\ }\bibfield  {title} {\bibinfo {title} {Quantum teleportation in high dimensions},\ }\href {https://doi.org/10.1103/PhysRevLett.123.070505} {\bibfield  {journal} {\bibinfo  {journal} {Phys. Rev. Lett.}\ }\textbf {\bibinfo {volume} {123}},\
  \bibinfo {pages} {070505} (\bibinfo {year} {2019})}\BibitemShut {NoStop}%
\bibitem [{\citenamefont {Collins}\ \emph {et~al.}(2002)\citenamefont {Collins}, \citenamefont {Gisin}, \citenamefont {Linden}, \citenamefont {Massar},\ and\ \citenamefont {Popescu}}]{collins2002BellInequalitiesArbitrarily}%
  \BibitemOpen
  \bibfield  {author} {\bibinfo {author} {\bibfnamefont {D.}~\bibnamefont {Collins}}, \bibinfo {author} {\bibfnamefont {N.}~\bibnamefont {Gisin}}, \bibinfo {author} {\bibfnamefont {N.}~\bibnamefont {Linden}}, \bibinfo {author} {\bibfnamefont {S.}~\bibnamefont {Massar}},\ and\ \bibinfo {author} {\bibfnamefont {S.}~\bibnamefont {Popescu}},\ }\bibfield  {title} {\bibinfo {title} {Bell {{Inequalities}} for {{Arbitrarily High-Dimensional Systems}}},\ }\href {https://doi.org/10.1103/PhysRevLett.88.040404} {\bibfield  {journal} {\bibinfo  {journal} {Phys. Rev. Lett.}\ }\textbf {\bibinfo {volume} {88}},\ \bibinfo {pages} {040404} (\bibinfo {year} {2002})}\BibitemShut {NoStop}%
\bibitem [{\citenamefont {Liu}\ \emph {et~al.}(2021)\citenamefont {Liu}, \citenamefont {Sun}, \citenamefont {Pachos}, \citenamefont {Yang}, \citenamefont {Meng}, \citenamefont {Liao}, \citenamefont {Li}, \citenamefont {Wang}, \citenamefont {Luo}, \citenamefont {He}, \citenamefont {Huang}, \citenamefont {Ding}, \citenamefont {Xu}, \citenamefont {Han}, \citenamefont {Li},\ and\ \citenamefont {Guo}}]{liu2021}%
  \BibitemOpen
  \bibfield  {author} {\bibinfo {author} {\bibfnamefont {Z.-H.}\ \bibnamefont {Liu}}, \bibinfo {author} {\bibfnamefont {K.}~\bibnamefont {Sun}}, \bibinfo {author} {\bibfnamefont {J.~K.}\ \bibnamefont {Pachos}}, \bibinfo {author} {\bibfnamefont {M.}~\bibnamefont {Yang}}, \bibinfo {author} {\bibfnamefont {Y.}~\bibnamefont {Meng}}, \bibinfo {author} {\bibfnamefont {Y.-W.}\ \bibnamefont {Liao}}, \bibinfo {author} {\bibfnamefont {Q.}~\bibnamefont {Li}}, \bibinfo {author} {\bibfnamefont {J.-F.}\ \bibnamefont {Wang}}, \bibinfo {author} {\bibfnamefont {Z.-Y.}\ \bibnamefont {Luo}}, \bibinfo {author} {\bibfnamefont {Y.-F.}\ \bibnamefont {He}}, \bibinfo {author} {\bibfnamefont {D.-Y.}\ \bibnamefont {Huang}}, \bibinfo {author} {\bibfnamefont {G.-R.}\ \bibnamefont {Ding}}, \bibinfo {author} {\bibfnamefont {J.-S.}\ \bibnamefont {Xu}}, \bibinfo {author} {\bibfnamefont {Y.-J.}\ \bibnamefont {Han}}, \bibinfo {author} {\bibfnamefont {C.-F.}\ \bibnamefont {Li}},\ and\ \bibinfo {author} {\bibfnamefont {G.-C.}\ \bibnamefont
  {Guo}},\ }\bibfield  {title} {\bibinfo {title} {Topological {{Contextuality}} and {{Anyonic Statistics}} of {{Photonic-Encoded Parafermions}}},\ }\href {https://doi.org/10.1103/PRXQuantum.2.030323} {\bibfield  {journal} {\bibinfo  {journal} {PRX Quantum}\ }\textbf {\bibinfo {volume} {2}},\ \bibinfo {pages} {030323} (\bibinfo {year} {2021})}\BibitemShut {NoStop}%
\bibitem [{\citenamefont {Jansen}\ \emph {et~al.}(2024)\citenamefont {Jansen}, \citenamefont {Heightman}, \citenamefont {Mortimer}, \citenamefont {Perito},\ and\ \citenamefont {Acín}}]{acin2024}%
  \BibitemOpen
  \bibfield  {author} {\bibinfo {author} {\bibfnamefont {D.}~\bibnamefont {Jansen}}, \bibinfo {author} {\bibfnamefont {T.}~\bibnamefont {Heightman}}, \bibinfo {author} {\bibfnamefont {L.}~\bibnamefont {Mortimer}}, \bibinfo {author} {\bibfnamefont {I.}~\bibnamefont {Perito}},\ and\ \bibinfo {author} {\bibfnamefont {A.}~\bibnamefont {Acín}},\ }\href {https://arxiv.org/abs/2406.00792} {\bibinfo {title} {Qudit inspired optimization for graph coloring}} (\bibinfo {year} {2024}),\ \Eprint {https://arxiv.org/abs/2406.00792} {arXiv:2406.00792 [quant-ph]} \BibitemShut {NoStop}%
\bibitem [{\citenamefont {Bienfait}\ \emph {et~al.}(2020)\citenamefont {Bienfait}, \citenamefont {Zhong}, \citenamefont {Chang}, \citenamefont {Chou}, \citenamefont {Conner}, \citenamefont {Dumur}, \citenamefont {Grebel}, \citenamefont {Peairs}, \citenamefont {Povey}, \citenamefont {Satzinger},\ and\ \citenamefont {Cleland}}]{Bienfait2020}%
  \BibitemOpen
  \bibfield  {author} {\bibinfo {author} {\bibfnamefont {A.}~\bibnamefont {Bienfait}}, \bibinfo {author} {\bibfnamefont {Y.~P.}\ \bibnamefont {Zhong}}, \bibinfo {author} {\bibfnamefont {H.-S.}\ \bibnamefont {Chang}}, \bibinfo {author} {\bibfnamefont {M.-H.}\ \bibnamefont {Chou}}, \bibinfo {author} {\bibfnamefont {C.~R.}\ \bibnamefont {Conner}}, \bibinfo {author} {\bibfnamefont {E.}~\bibnamefont {Dumur}}, \bibinfo {author} {\bibfnamefont {J.}~\bibnamefont {Grebel}}, \bibinfo {author} {\bibfnamefont {G.~A.}\ \bibnamefont {Peairs}}, \bibinfo {author} {\bibfnamefont {R.~G.}\ \bibnamefont {Povey}}, \bibinfo {author} {\bibfnamefont {K.~J.}\ \bibnamefont {Satzinger}},\ and\ \bibinfo {author} {\bibfnamefont {A.~N.}\ \bibnamefont {Cleland}},\ }\bibfield  {title} {\bibinfo {title} {Quantum erasure using entangled surface acoustic phonons},\ }\href {https://doi.org/10.1103/PhysRevX.10.021055} {\bibfield  {journal} {\bibinfo  {journal} {Phys. Rev. X}\ }\textbf {\bibinfo {volume} {10}},\ \bibinfo {pages} {021055}
  (\bibinfo {year} {2020})}\BibitemShut {NoStop}%
\bibitem [{\citenamefont {Kim}\ \emph {et~al.}(2024)\citenamefont {Kim}, \citenamefont {Park}, \citenamefont {Baek}, \citenamefont {Lee},\ and\ \citenamefont {Moon}}]{kim2024}%
  \BibitemOpen
  \bibfield  {author} {\bibinfo {author} {\bibfnamefont {D.}~\bibnamefont {Kim}}, \bibinfo {author} {\bibfnamefont {J.}~\bibnamefont {Park}}, \bibinfo {author} {\bibfnamefont {C.}~\bibnamefont {Baek}}, \bibinfo {author} {\bibfnamefont {S.~K.}\ \bibnamefont {Lee}},\ and\ \bibinfo {author} {\bibfnamefont {H.~S.}\ \bibnamefont {Moon}},\ }\bibfield  {title} {\bibinfo {title} {Complementarity of which-path information in induced and stimulated coherences via four-wave mixing process from warm {{Rb}} atomic ensemble},\ }\href {https://doi.org/10.1364/OPTICAQ.528135} {\bibfield  {journal} {\bibinfo  {journal} {Optica Quantum}\ }\textbf {\bibinfo {volume} {2}},\ \bibinfo {pages} {288} (\bibinfo {year} {2024})}\BibitemShut {NoStop}%
\bibitem [{\citenamefont {Nisbet-Jones}\ \emph {et~al.}(2013)\citenamefont {Nisbet-Jones}, \citenamefont {Dilley}, \citenamefont {Holleczek}, \citenamefont {Barter},\ and\ \citenamefont {Kuhn}}]{NisbetJones2013}%
  \BibitemOpen
  \bibfield  {author} {\bibinfo {author} {\bibfnamefont {P.~B.~R.}\ \bibnamefont {Nisbet-Jones}}, \bibinfo {author} {\bibfnamefont {J.}~\bibnamefont {Dilley}}, \bibinfo {author} {\bibfnamefont {A.}~\bibnamefont {Holleczek}}, \bibinfo {author} {\bibfnamefont {O.}~\bibnamefont {Barter}},\ and\ \bibinfo {author} {\bibfnamefont {A.}~\bibnamefont {Kuhn}},\ }\bibfield  {title} {\bibinfo {title} {Photonic qubits, qutrits and ququads accurately prepared and delivered on demand},\ }\href {https://doi.org/10.1088/1367-2630/15/5/053007} {\bibfield  {journal} {\bibinfo  {journal} {New J. Phys.}\ }\textbf {\bibinfo {volume} {15}},\ \bibinfo {pages} {053007} (\bibinfo {year} {2013})}\BibitemShut {NoStop}%
\bibitem [{\citenamefont {Blok}\ \emph {et~al.}(2021{\natexlab{b}})\citenamefont {Blok}, \citenamefont {Ramasesh}, \citenamefont {Schuster}, \citenamefont {O'Brien}, \citenamefont {Kreikebaum}, \citenamefont {Dahlen}, \citenamefont {Morvan}, \citenamefont {Yoshida}, \citenamefont {Yao},\ and\ \citenamefont {Siddiqi}}]{Qutrit2021}%
  \BibitemOpen
  \bibfield  {author} {\bibinfo {author} {\bibfnamefont {M.~S.}\ \bibnamefont {Blok}}, \bibinfo {author} {\bibfnamefont {V.~V.}\ \bibnamefont {Ramasesh}}, \bibinfo {author} {\bibfnamefont {T.}~\bibnamefont {Schuster}}, \bibinfo {author} {\bibfnamefont {K.}~\bibnamefont {O'Brien}}, \bibinfo {author} {\bibfnamefont {J.~M.}\ \bibnamefont {Kreikebaum}}, \bibinfo {author} {\bibfnamefont {D.}~\bibnamefont {Dahlen}}, \bibinfo {author} {\bibfnamefont {A.}~\bibnamefont {Morvan}}, \bibinfo {author} {\bibfnamefont {B.}~\bibnamefont {Yoshida}}, \bibinfo {author} {\bibfnamefont {N.~Y.}\ \bibnamefont {Yao}},\ and\ \bibinfo {author} {\bibfnamefont {I.}~\bibnamefont {Siddiqi}},\ }\bibfield  {title} {\bibinfo {title} {Quantum information scrambling on a superconducting qutrit processor},\ }\href {https://doi.org/10.1103/PhysRevX.11.021010} {\bibfield  {journal} {\bibinfo  {journal} {Phys. Rev. X}\ }\textbf {\bibinfo {volume} {11}},\ \bibinfo {pages} {021010} (\bibinfo {year} {2021}{\natexlab{b}})}\BibitemShut {NoStop}%
\bibitem [{\citenamefont {Goss}\ \emph {et~al.}(2022)\citenamefont {Goss}, \citenamefont {Morvan}, \citenamefont {Marinelli}, \citenamefont {Mitchell}, \citenamefont {Nguyen}, \citenamefont {Naik}, \citenamefont {Chen}, \citenamefont {J{\"u}nger}, \citenamefont {Kreikebaum}, \citenamefont {Santiago}, \citenamefont {Wallman},\ and\ \citenamefont {Siddiqi}}]{Goss2022}%
  \BibitemOpen
  \bibfield  {author} {\bibinfo {author} {\bibfnamefont {N.}~\bibnamefont {Goss}}, \bibinfo {author} {\bibfnamefont {A.}~\bibnamefont {Morvan}}, \bibinfo {author} {\bibfnamefont {B.}~\bibnamefont {Marinelli}}, \bibinfo {author} {\bibfnamefont {B.~K.}\ \bibnamefont {Mitchell}}, \bibinfo {author} {\bibfnamefont {L.~B.}\ \bibnamefont {Nguyen}}, \bibinfo {author} {\bibfnamefont {R.~K.}\ \bibnamefont {Naik}}, \bibinfo {author} {\bibfnamefont {L.}~\bibnamefont {Chen}}, \bibinfo {author} {\bibfnamefont {C.}~\bibnamefont {J{\"u}nger}}, \bibinfo {author} {\bibfnamefont {J.~M.}\ \bibnamefont {Kreikebaum}}, \bibinfo {author} {\bibfnamefont {D.~I.}\ \bibnamefont {Santiago}}, \bibinfo {author} {\bibfnamefont {J.~J.}\ \bibnamefont {Wallman}},\ and\ \bibinfo {author} {\bibfnamefont {I.}~\bibnamefont {Siddiqi}},\ }\bibfield  {title} {\bibinfo {title} {High-fidelity qutrit entangling gates for superconducting circuits},\ }\href {https://doi.org/10.1038/s41467-022-34851-z} {\bibfield  {journal} {\bibinfo  {journal} {Nat.
  Commun.}\ }\textbf {\bibinfo {volume} {13}},\ \bibinfo {pages} {7481} (\bibinfo {year} {2022})}\BibitemShut {NoStop}%
\bibitem [{\citenamefont {Luo}\ \emph {et~al.}(2023)\citenamefont {Luo}, \citenamefont {Huang}, \citenamefont {Tao}, \citenamefont {Zhang}, \citenamefont {Zhou}, \citenamefont {Chu}, \citenamefont {Liu}, \citenamefont {Wang}, \citenamefont {Cui}, \citenamefont {Liu}, \citenamefont {Yan}, \citenamefont {Yung}, \citenamefont {Chen}, \citenamefont {Yan},\ and\ \citenamefont {Yu}}]{Superconducting2023PRL}%
  \BibitemOpen
  \bibfield  {author} {\bibinfo {author} {\bibfnamefont {K.}~\bibnamefont {Luo}}, \bibinfo {author} {\bibfnamefont {W.}~\bibnamefont {Huang}}, \bibinfo {author} {\bibfnamefont {Z.}~\bibnamefont {Tao}}, \bibinfo {author} {\bibfnamefont {L.}~\bibnamefont {Zhang}}, \bibinfo {author} {\bibfnamefont {Y.}~\bibnamefont {Zhou}}, \bibinfo {author} {\bibfnamefont {J.}~\bibnamefont {Chu}}, \bibinfo {author} {\bibfnamefont {W.}~\bibnamefont {Liu}}, \bibinfo {author} {\bibfnamefont {B.}~\bibnamefont {Wang}}, \bibinfo {author} {\bibfnamefont {J.}~\bibnamefont {Cui}}, \bibinfo {author} {\bibfnamefont {S.}~\bibnamefont {Liu}}, \bibinfo {author} {\bibfnamefont {F.}~\bibnamefont {Yan}}, \bibinfo {author} {\bibfnamefont {M.-H.}\ \bibnamefont {Yung}}, \bibinfo {author} {\bibfnamefont {Y.}~\bibnamefont {Chen}}, \bibinfo {author} {\bibfnamefont {T.}~\bibnamefont {Yan}},\ and\ \bibinfo {author} {\bibfnamefont {D.}~\bibnamefont {Yu}},\ }\bibfield  {title} {\bibinfo {title} {Experimental realization of two qutrits gate with tunable
  coupling in superconducting circuits},\ }\href {https://doi.org/10.1103/PhysRevLett.130.030603} {\bibfield  {journal} {\bibinfo  {journal} {Phys. Rev. Lett.}\ }\textbf {\bibinfo {volume} {130}},\ \bibinfo {pages} {030603} (\bibinfo {year} {2023})}\BibitemShut {NoStop}%
\bibitem [{\citenamefont {Klimov}\ \emph {et~al.}(2003)\citenamefont {Klimov}, \citenamefont {Guzm\'an}, \citenamefont {Retamal},\ and\ \citenamefont {Saavedra}}]{Qutrit_ion_2003}%
  \BibitemOpen
  \bibfield  {author} {\bibinfo {author} {\bibfnamefont {A.~B.}\ \bibnamefont {Klimov}}, \bibinfo {author} {\bibfnamefont {R.}~\bibnamefont {Guzm\'an}}, \bibinfo {author} {\bibfnamefont {J.~C.}\ \bibnamefont {Retamal}},\ and\ \bibinfo {author} {\bibfnamefont {C.}~\bibnamefont {Saavedra}},\ }\bibfield  {title} {\bibinfo {title} {Qutrit quantum computer with trapped ions},\ }\href {https://doi.org/10.1103/PhysRevA.67.062313} {\bibfield  {journal} {\bibinfo  {journal} {Phys. Rev. A}\ }\textbf {\bibinfo {volume} {67}},\ \bibinfo {pages} {062313} (\bibinfo {year} {2003})}\BibitemShut {NoStop}%
\bibitem [{\citenamefont {Low}\ \emph {et~al.}(2020)\citenamefont {Low}, \citenamefont {White}, \citenamefont {Cox}, \citenamefont {Day},\ and\ \citenamefont {Senko}}]{Practical_PRR_2020}%
  \BibitemOpen
  \bibfield  {author} {\bibinfo {author} {\bibfnamefont {P.~J.}\ \bibnamefont {Low}}, \bibinfo {author} {\bibfnamefont {B.~M.}\ \bibnamefont {White}}, \bibinfo {author} {\bibfnamefont {A.~A.}\ \bibnamefont {Cox}}, \bibinfo {author} {\bibfnamefont {M.~L.}\ \bibnamefont {Day}},\ and\ \bibinfo {author} {\bibfnamefont {C.}~\bibnamefont {Senko}},\ }\bibfield  {title} {\bibinfo {title} {Practical trapped-ion protocols for universal qudit-based quantum computing},\ }\href {https://doi.org/10.1103/PhysRevResearch.2.033128} {\bibfield  {journal} {\bibinfo  {journal} {Phys. Rev. Res.}\ }\textbf {\bibinfo {volume} {2}},\ \bibinfo {pages} {033128} (\bibinfo {year} {2020})}\BibitemShut {NoStop}%
\bibitem [{\citenamefont {Ringbauer}\ \emph {et~al.}(2022)\citenamefont {Ringbauer}, \citenamefont {Meth}, \citenamefont {Postler}, \citenamefont {Stricker}, \citenamefont {Blatt}, \citenamefont {Schindler},\ and\ \citenamefont {Monz}}]{ion_Ringbauer2022}%
  \BibitemOpen
  \bibfield  {author} {\bibinfo {author} {\bibfnamefont {M.}~\bibnamefont {Ringbauer}}, \bibinfo {author} {\bibfnamefont {M.}~\bibnamefont {Meth}}, \bibinfo {author} {\bibfnamefont {L.}~\bibnamefont {Postler}}, \bibinfo {author} {\bibfnamefont {R.}~\bibnamefont {Stricker}}, \bibinfo {author} {\bibfnamefont {R.}~\bibnamefont {Blatt}}, \bibinfo {author} {\bibfnamefont {P.}~\bibnamefont {Schindler}},\ and\ \bibinfo {author} {\bibfnamefont {T.}~\bibnamefont {Monz}},\ }\bibfield  {title} {\bibinfo {title} {A universal qudit quantum processor with trapped ions},\ }\href {https://doi.org/10.1038/s41567-022-01658-0} {\bibfield  {journal} {\bibinfo  {journal} {Nat. Phys.}\ }\textbf {\bibinfo {volume} {18}},\ \bibinfo {pages} {1053} (\bibinfo {year} {2022})}\BibitemShut {NoStop}%
\bibitem [{\citenamefont {Hrmo}\ \emph {et~al.}(2023)\citenamefont {Hrmo}, \citenamefont {Wilhelm}, \citenamefont {Gerster}, \citenamefont {van Mourik}, \citenamefont {Huber}, \citenamefont {Blatt}, \citenamefont {Schindler}, \citenamefont {Monz},\ and\ \citenamefont {Ringbauer}}]{Hrmo2023}%
  \BibitemOpen
  \bibfield  {author} {\bibinfo {author} {\bibfnamefont {P.}~\bibnamefont {Hrmo}}, \bibinfo {author} {\bibfnamefont {B.}~\bibnamefont {Wilhelm}}, \bibinfo {author} {\bibfnamefont {L.}~\bibnamefont {Gerster}}, \bibinfo {author} {\bibfnamefont {M.~W.}\ \bibnamefont {van Mourik}}, \bibinfo {author} {\bibfnamefont {M.}~\bibnamefont {Huber}}, \bibinfo {author} {\bibfnamefont {R.}~\bibnamefont {Blatt}}, \bibinfo {author} {\bibfnamefont {P.}~\bibnamefont {Schindler}}, \bibinfo {author} {\bibfnamefont {T.}~\bibnamefont {Monz}},\ and\ \bibinfo {author} {\bibfnamefont {M.}~\bibnamefont {Ringbauer}},\ }\bibfield  {title} {\bibinfo {title} {Native qudit entanglement in a trapped ion quantum processor},\ }\href {https://doi.org/10.1038/s41467-023-37375-2} {\bibfield  {journal} {\bibinfo  {journal} {Nat. Commun.}\ }\textbf {\bibinfo {volume} {14}},\ \bibinfo {pages} {2242} (\bibinfo {year} {2023})}\BibitemShut {NoStop}%
\bibitem [{\citenamefont {Saffman}\ \emph {et~al.}(2010)\citenamefont {Saffman}, \citenamefont {Walker},\ and\ \citenamefont {M\o{}lmer}}]{Saffman2010}%
  \BibitemOpen
  \bibfield  {author} {\bibinfo {author} {\bibfnamefont {M.}~\bibnamefont {Saffman}}, \bibinfo {author} {\bibfnamefont {T.~G.}\ \bibnamefont {Walker}},\ and\ \bibinfo {author} {\bibfnamefont {K.}~\bibnamefont {M\o{}lmer}},\ }\bibfield  {title} {\bibinfo {title} {Quantum information with rydberg atoms},\ }\href {https://doi.org/10.1103/RevModPhys.82.2313} {\bibfield  {journal} {\bibinfo  {journal} {Rev. Mod. Phys.}\ }\textbf {\bibinfo {volume} {82}},\ \bibinfo {pages} {2313} (\bibinfo {year} {2010})}\BibitemShut {NoStop}%
\bibitem [{\citenamefont {Bernien}\ \emph {et~al.}(2017)\citenamefont {Bernien}, \citenamefont {Schwartz}, \citenamefont {Keesling}, \citenamefont {Levine}, \citenamefont {Omran}, \citenamefont {Pichler}, \citenamefont {Choi}, \citenamefont {Zibrov}, \citenamefont {Endres}, \citenamefont {Greiner}, \citenamefont {Vuleti{\'c}},\ and\ \citenamefont {Lukin}}]{bernienProbingManybodyDynamics2017}%
  \BibitemOpen
  \bibfield  {author} {\bibinfo {author} {\bibfnamefont {H.}~\bibnamefont {Bernien}}, \bibinfo {author} {\bibfnamefont {S.}~\bibnamefont {Schwartz}}, \bibinfo {author} {\bibfnamefont {A.}~\bibnamefont {Keesling}}, \bibinfo {author} {\bibfnamefont {H.}~\bibnamefont {Levine}}, \bibinfo {author} {\bibfnamefont {A.}~\bibnamefont {Omran}}, \bibinfo {author} {\bibfnamefont {H.}~\bibnamefont {Pichler}}, \bibinfo {author} {\bibfnamefont {S.}~\bibnamefont {Choi}}, \bibinfo {author} {\bibfnamefont {A.~S.}\ \bibnamefont {Zibrov}}, \bibinfo {author} {\bibfnamefont {M.}~\bibnamefont {Endres}}, \bibinfo {author} {\bibfnamefont {M.}~\bibnamefont {Greiner}}, \bibinfo {author} {\bibfnamefont {V.}~\bibnamefont {Vuleti{\'c}}},\ and\ \bibinfo {author} {\bibfnamefont {M.~D.}\ \bibnamefont {Lukin}},\ }\bibfield  {title} {\bibinfo {title} {Probing many-body dynamics on a 51-atom quantum simulator},\ }\href {https://doi.org/10.1038/nature24622} {\bibfield  {journal} {\bibinfo  {journal} {Nature}\ }\textbf {\bibinfo {volume} {551}},\
  \bibinfo {pages} {579} (\bibinfo {year} {2017})}\BibitemShut {NoStop}%
\bibitem [{\citenamefont {de~L{\'e}s{\'e}leuc}\ \emph {et~al.}(2019)\citenamefont {de~L{\'e}s{\'e}leuc}, \citenamefont {Lienhard}, \citenamefont {Scholl}, \citenamefont {Barredo}, \citenamefont {Weber}, \citenamefont {Lang}, \citenamefont {B{\"u}chler}, \citenamefont {Lahaye},\ and\ \citenamefont {Browaeys}}]{leseleucObservationSymmetryprotectedTopological2019}%
  \BibitemOpen
  \bibfield  {author} {\bibinfo {author} {\bibfnamefont {S.}~\bibnamefont {de~L{\'e}s{\'e}leuc}}, \bibinfo {author} {\bibfnamefont {V.}~\bibnamefont {Lienhard}}, \bibinfo {author} {\bibfnamefont {P.}~\bibnamefont {Scholl}}, \bibinfo {author} {\bibfnamefont {D.}~\bibnamefont {Barredo}}, \bibinfo {author} {\bibfnamefont {S.}~\bibnamefont {Weber}}, \bibinfo {author} {\bibfnamefont {N.}~\bibnamefont {Lang}}, \bibinfo {author} {\bibfnamefont {H.~P.}\ \bibnamefont {B{\"u}chler}}, \bibinfo {author} {\bibfnamefont {T.}~\bibnamefont {Lahaye}},\ and\ \bibinfo {author} {\bibfnamefont {A.}~\bibnamefont {Browaeys}},\ }\bibfield  {title} {\bibinfo {title} {Observation of a symmetry-protected topological phase of interacting bosons with {{Rydberg}} atoms},\ }\href {https://doi.org/10.1126/science.aav9105} {\bibfield  {journal} {\bibinfo  {journal} {Science}\ }\textbf {\bibinfo {volume} {365}},\ \bibinfo {pages} {775} (\bibinfo {year} {2019})}\BibitemShut {NoStop}%
\bibitem [{\citenamefont {Scholl}\ \emph {et~al.}(2021)\citenamefont {Scholl}, \citenamefont {Schuler}, \citenamefont {Williams}, \citenamefont {Eberharter}, \citenamefont {Barredo}, \citenamefont {Schymik}, \citenamefont {Lienhard}, \citenamefont {Henry}, \citenamefont {Lang}, \citenamefont {Lahaye}, \citenamefont {L{\"a}uchli},\ and\ \citenamefont {Browaeys}}]{schollQuantumSimulation2D2021}%
  \BibitemOpen
  \bibfield  {author} {\bibinfo {author} {\bibfnamefont {P.}~\bibnamefont {Scholl}}, \bibinfo {author} {\bibfnamefont {M.}~\bibnamefont {Schuler}}, \bibinfo {author} {\bibfnamefont {H.~J.}\ \bibnamefont {Williams}}, \bibinfo {author} {\bibfnamefont {A.~A.}\ \bibnamefont {Eberharter}}, \bibinfo {author} {\bibfnamefont {D.}~\bibnamefont {Barredo}}, \bibinfo {author} {\bibfnamefont {K.-N.}\ \bibnamefont {Schymik}}, \bibinfo {author} {\bibfnamefont {V.}~\bibnamefont {Lienhard}}, \bibinfo {author} {\bibfnamefont {L.-P.}\ \bibnamefont {Henry}}, \bibinfo {author} {\bibfnamefont {T.~C.}\ \bibnamefont {Lang}}, \bibinfo {author} {\bibfnamefont {T.}~\bibnamefont {Lahaye}}, \bibinfo {author} {\bibfnamefont {A.~M.}\ \bibnamefont {L{\"a}uchli}},\ and\ \bibinfo {author} {\bibfnamefont {A.}~\bibnamefont {Browaeys}},\ }\bibfield  {title} {\bibinfo {title} {Quantum simulation of {{2D}} antiferromagnets with hundreds of {{Rydberg}} atoms},\ }\href {https://doi.org/10.1038/s41586-021-03585-1} {\bibfield  {journal} {\bibinfo
  {journal} {Nature}\ }\textbf {\bibinfo {volume} {595}},\ \bibinfo {pages} {233} (\bibinfo {year} {2021})}\BibitemShut {NoStop}%
\bibitem [{\citenamefont {Ebadi}\ \emph {et~al.}(2021)\citenamefont {Ebadi}, \citenamefont {Wang}, \citenamefont {Levine}, \citenamefont {Keesling}, \citenamefont {Semeghini}, \citenamefont {Omran}, \citenamefont {Bluvstein}, \citenamefont {Samajdar}, \citenamefont {Pichler}, \citenamefont {Ho}, \citenamefont {Choi}, \citenamefont {Sachdev}, \citenamefont {Greiner}, \citenamefont {Vuleti{\'c}},\ and\ \citenamefont {Lukin}}]{ebadiQuantumPhasesMatter2021}%
  \BibitemOpen
  \bibfield  {author} {\bibinfo {author} {\bibfnamefont {S.}~\bibnamefont {Ebadi}}, \bibinfo {author} {\bibfnamefont {T.~T.}\ \bibnamefont {Wang}}, \bibinfo {author} {\bibfnamefont {H.}~\bibnamefont {Levine}}, \bibinfo {author} {\bibfnamefont {A.}~\bibnamefont {Keesling}}, \bibinfo {author} {\bibfnamefont {G.}~\bibnamefont {Semeghini}}, \bibinfo {author} {\bibfnamefont {A.}~\bibnamefont {Omran}}, \bibinfo {author} {\bibfnamefont {D.}~\bibnamefont {Bluvstein}}, \bibinfo {author} {\bibfnamefont {R.}~\bibnamefont {Samajdar}}, \bibinfo {author} {\bibfnamefont {H.}~\bibnamefont {Pichler}}, \bibinfo {author} {\bibfnamefont {W.~W.}\ \bibnamefont {Ho}}, \bibinfo {author} {\bibfnamefont {S.}~\bibnamefont {Choi}}, \bibinfo {author} {\bibfnamefont {S.}~\bibnamefont {Sachdev}}, \bibinfo {author} {\bibfnamefont {M.}~\bibnamefont {Greiner}}, \bibinfo {author} {\bibfnamefont {V.}~\bibnamefont {Vuleti{\'c}}},\ and\ \bibinfo {author} {\bibfnamefont {M.~D.}\ \bibnamefont {Lukin}},\ }\bibfield  {title} {\bibinfo {title}
  {Quantum phases of matter on a 256-atom programmable quantum simulator},\ }\href {https://doi.org/10.1038/s41586-021-03582-4} {\bibfield  {journal} {\bibinfo  {journal} {Nature}\ }\textbf {\bibinfo {volume} {595}},\ \bibinfo {pages} {227} (\bibinfo {year} {2021})}\BibitemShut {NoStop}%
\bibitem [{\citenamefont {Bornet}\ \emph {et~al.}(2023)\citenamefont {Bornet}, \citenamefont {Emperauger}, \citenamefont {Chen}, \citenamefont {Ye}, \citenamefont {Block}, \citenamefont {Bintz}, \citenamefont {Boyd}, \citenamefont {Barredo}, \citenamefont {Comparin}, \citenamefont {Mezzacapo}, \citenamefont {Roscilde}, \citenamefont {Lahaye}, \citenamefont {Yao},\ and\ \citenamefont {Browaeys}}]{bornet2023}%
  \BibitemOpen
  \bibfield  {author} {\bibinfo {author} {\bibfnamefont {G.}~\bibnamefont {Bornet}}, \bibinfo {author} {\bibfnamefont {G.}~\bibnamefont {Emperauger}}, \bibinfo {author} {\bibfnamefont {C.}~\bibnamefont {Chen}}, \bibinfo {author} {\bibfnamefont {B.}~\bibnamefont {Ye}}, \bibinfo {author} {\bibfnamefont {M.}~\bibnamefont {Block}}, \bibinfo {author} {\bibfnamefont {M.}~\bibnamefont {Bintz}}, \bibinfo {author} {\bibfnamefont {J.~A.}\ \bibnamefont {Boyd}}, \bibinfo {author} {\bibfnamefont {D.}~\bibnamefont {Barredo}}, \bibinfo {author} {\bibfnamefont {T.}~\bibnamefont {Comparin}}, \bibinfo {author} {\bibfnamefont {F.}~\bibnamefont {Mezzacapo}}, \bibinfo {author} {\bibfnamefont {T.}~\bibnamefont {Roscilde}}, \bibinfo {author} {\bibfnamefont {T.}~\bibnamefont {Lahaye}}, \bibinfo {author} {\bibfnamefont {N.~Y.}\ \bibnamefont {Yao}},\ and\ \bibinfo {author} {\bibfnamefont {A.}~\bibnamefont {Browaeys}},\ }\bibfield  {title} {\bibinfo {title} {Scalable spin squeezing in a dipolar {{Rydberg}} atom array},\ }\href
  {https://doi.org/10.1038/s41586-023-06414-9} {\bibfield  {journal} {\bibinfo  {journal} {Nature}\ }\textbf {\bibinfo {volume} {621}},\ \bibinfo {pages} {728} (\bibinfo {year} {2023})}\BibitemShut {NoStop}%
\bibitem [{\citenamefont {Bluvstein}\ \emph {et~al.}(2024)\citenamefont {Bluvstein}, \citenamefont {Evered}, \citenamefont {Geim}, \citenamefont {Li}, \citenamefont {Zhou}, \citenamefont {Manovitz}, \citenamefont {Ebadi}, \citenamefont {Cain}, \citenamefont {Kalinowski}, \citenamefont {Hangleiter}, \citenamefont {Bonilla~Ataides}, \citenamefont {Maskara}, \citenamefont {Cong}, \citenamefont {Gao}, \citenamefont {Sales~Rodriguez}, \citenamefont {Karolyshyn}, \citenamefont {Semeghini}, \citenamefont {Gullans}, \citenamefont {Greiner}, \citenamefont {Vuleti{\'c}},\ and\ \citenamefont {Lukin}}]{bluvstein2024}%
  \BibitemOpen
  \bibfield  {author} {\bibinfo {author} {\bibfnamefont {D.}~\bibnamefont {Bluvstein}}, \bibinfo {author} {\bibfnamefont {S.~J.}\ \bibnamefont {Evered}}, \bibinfo {author} {\bibfnamefont {A.~A.}\ \bibnamefont {Geim}}, \bibinfo {author} {\bibfnamefont {S.~H.}\ \bibnamefont {Li}}, \bibinfo {author} {\bibfnamefont {H.}~\bibnamefont {Zhou}}, \bibinfo {author} {\bibfnamefont {T.}~\bibnamefont {Manovitz}}, \bibinfo {author} {\bibfnamefont {S.}~\bibnamefont {Ebadi}}, \bibinfo {author} {\bibfnamefont {M.}~\bibnamefont {Cain}}, \bibinfo {author} {\bibfnamefont {M.}~\bibnamefont {Kalinowski}}, \bibinfo {author} {\bibfnamefont {D.}~\bibnamefont {Hangleiter}}, \bibinfo {author} {\bibfnamefont {J.~P.}\ \bibnamefont {Bonilla~Ataides}}, \bibinfo {author} {\bibfnamefont {N.}~\bibnamefont {Maskara}}, \bibinfo {author} {\bibfnamefont {I.}~\bibnamefont {Cong}}, \bibinfo {author} {\bibfnamefont {X.}~\bibnamefont {Gao}}, \bibinfo {author} {\bibfnamefont {P.}~\bibnamefont {Sales~Rodriguez}}, \bibinfo {author} {\bibfnamefont
  {T.}~\bibnamefont {Karolyshyn}}, \bibinfo {author} {\bibfnamefont {G.}~\bibnamefont {Semeghini}}, \bibinfo {author} {\bibfnamefont {M.~J.}\ \bibnamefont {Gullans}}, \bibinfo {author} {\bibfnamefont {M.}~\bibnamefont {Greiner}}, \bibinfo {author} {\bibfnamefont {V.}~\bibnamefont {Vuleti{\'c}}},\ and\ \bibinfo {author} {\bibfnamefont {M.~D.}\ \bibnamefont {Lukin}},\ }\bibfield  {title} {\bibinfo {title} {Logical quantum processor based on reconfigurable atom arrays},\ }\href {https://doi.org/10.1038/s41586-023-06927-3} {\bibfield  {journal} {\bibinfo  {journal} {Nature}\ }\textbf {\bibinfo {volume} {626}},\ \bibinfo {pages} {58} (\bibinfo {year} {2024})}\BibitemShut {NoStop}%
\bibitem [{\citenamefont {Brion}\ \emph {et~al.}(2007)\citenamefont {Brion}, \citenamefont {M{\o}lmer},\ and\ \citenamefont {Saffman}}]{brionQuantumComputingCollective2007}%
  \BibitemOpen
  \bibfield  {author} {\bibinfo {author} {\bibfnamefont {E.}~\bibnamefont {Brion}}, \bibinfo {author} {\bibfnamefont {K.}~\bibnamefont {M{\o}lmer}},\ and\ \bibinfo {author} {\bibfnamefont {M.}~\bibnamefont {Saffman}},\ }\bibfield  {title} {\bibinfo {title} {Quantum {{Computing}} with {{Collective Ensembles}} of {{Multilevel Systems}}},\ }\href {https://doi.org/10.1103/PhysRevLett.99.260501} {\bibfield  {journal} {\bibinfo  {journal} {Phys. Rev. Lett.}\ }\textbf {\bibinfo {volume} {99}},\ \bibinfo {pages} {260501} (\bibinfo {year} {2007})}\BibitemShut {NoStop}%
\bibitem [{\citenamefont {Lukin}\ \emph {et~al.}(2001)\citenamefont {Lukin}, \citenamefont {Fleischhauer}, \citenamefont {Cote}, \citenamefont {Duan}, \citenamefont {Jaksch}, \citenamefont {Cirac},\ and\ \citenamefont {Zoller}}]{Dipole_Lukin_2001}%
  \BibitemOpen
  \bibfield  {author} {\bibinfo {author} {\bibfnamefont {M.~D.}\ \bibnamefont {Lukin}}, \bibinfo {author} {\bibfnamefont {M.}~\bibnamefont {Fleischhauer}}, \bibinfo {author} {\bibfnamefont {R.}~\bibnamefont {Cote}}, \bibinfo {author} {\bibfnamefont {L.~M.}\ \bibnamefont {Duan}}, \bibinfo {author} {\bibfnamefont {D.}~\bibnamefont {Jaksch}}, \bibinfo {author} {\bibfnamefont {J.~I.}\ \bibnamefont {Cirac}},\ and\ \bibinfo {author} {\bibfnamefont {P.}~\bibnamefont {Zoller}},\ }\bibfield  {title} {\bibinfo {title} {Dipole blockade and quantum information processing in mesoscopic atomic ensembles},\ }\href {https://doi.org/10.1103/PhysRevLett.87.037901} {\bibfield  {journal} {\bibinfo  {journal} {Phys. Rev. Lett.}\ }\textbf {\bibinfo {volume} {87}},\ \bibinfo {pages} {037901} (\bibinfo {year} {2001})}\BibitemShut {NoStop}%
\bibitem [{\citenamefont {Bariani}\ \emph {et~al.}(2012)\citenamefont {Bariani}, \citenamefont {Dudin}, \citenamefont {Kennedy},\ and\ \citenamefont {Kuzmich}}]{Dephasing_Bariani_2012}%
  \BibitemOpen
  \bibfield  {author} {\bibinfo {author} {\bibfnamefont {F.}~\bibnamefont {Bariani}}, \bibinfo {author} {\bibfnamefont {Y.~O.}\ \bibnamefont {Dudin}}, \bibinfo {author} {\bibfnamefont {T.~A.~B.}\ \bibnamefont {Kennedy}},\ and\ \bibinfo {author} {\bibfnamefont {A.}~\bibnamefont {Kuzmich}},\ }\bibfield  {title} {\bibinfo {title} {Dephasing of multiparticle rydberg excitations for fast entanglement generation},\ }\href {https://doi.org/10.1103/PhysRevLett.108.030501} {\bibfield  {journal} {\bibinfo  {journal} {Phys. Rev. Lett.}\ }\textbf {\bibinfo {volume} {108}},\ \bibinfo {pages} {030501} (\bibinfo {year} {2012})}\BibitemShut {NoStop}%
\bibitem [{\citenamefont {Tresp}\ \emph {et~al.}(2015)\citenamefont {Tresp}, \citenamefont {Bienias}, \citenamefont {Weber}, \citenamefont {Gorniaczyk}, \citenamefont {Mirgorodskiy}, \citenamefont {B\"uchler},\ and\ \citenamefont {Hofferberth}}]{Dipolar_Tresp_2015}%
  \BibitemOpen
  \bibfield  {author} {\bibinfo {author} {\bibfnamefont {C.}~\bibnamefont {Tresp}}, \bibinfo {author} {\bibfnamefont {P.}~\bibnamefont {Bienias}}, \bibinfo {author} {\bibfnamefont {S.}~\bibnamefont {Weber}}, \bibinfo {author} {\bibfnamefont {H.}~\bibnamefont {Gorniaczyk}}, \bibinfo {author} {\bibfnamefont {I.}~\bibnamefont {Mirgorodskiy}}, \bibinfo {author} {\bibfnamefont {H.~P.}\ \bibnamefont {B\"uchler}},\ and\ \bibinfo {author} {\bibfnamefont {S.}~\bibnamefont {Hofferberth}},\ }\bibfield  {title} {\bibinfo {title} {Dipolar dephasing of rydberg $d$-state polaritons},\ }\href {https://doi.org/10.1103/PhysRevLett.115.083602} {\bibfield  {journal} {\bibinfo  {journal} {Phys. Rev. Lett.}\ }\textbf {\bibinfo {volume} {115}},\ \bibinfo {pages} {083602} (\bibinfo {year} {2015})}\BibitemShut {NoStop}%
\bibitem [{\citenamefont {Khazali}\ and\ \citenamefont {M{\o}lmer}(2020)}]{khazaliFastMultiqubitGates2020}%
  \BibitemOpen
  \bibfield  {author} {\bibinfo {author} {\bibfnamefont {M.}~\bibnamefont {Khazali}}\ and\ \bibinfo {author} {\bibfnamefont {K.}~\bibnamefont {M{\o}lmer}},\ }\bibfield  {title} {\bibinfo {title} {Fast {{Multiqubit Gates}} by {{Adiabatic Evolution}} in {{Interacting Excited-State Manifolds}} of {{Rydberg Atoms}} and {{Superconducting Circuits}}},\ }\href {https://doi.org/10.1103/PhysRevX.10.021054} {\bibfield  {journal} {\bibinfo  {journal} {Phys. Rev. X}\ }\textbf {\bibinfo {volume} {10}},\ \bibinfo {pages} {021054} (\bibinfo {year} {2020})}\BibitemShut {NoStop}%
\bibitem [{\citenamefont {Wu}\ \emph {et~al.}(2021)\citenamefont {Wu}, \citenamefont {Wang}, \citenamefont {Han}, \citenamefont {Jiang}, \citenamefont {Song}, \citenamefont {Xia}, \citenamefont {Su},\ and\ \citenamefont {Li}}]{wuSystematicErrorTolerantMultiqubitHolonomic2021}%
  \BibitemOpen
  \bibfield  {author} {\bibinfo {author} {\bibfnamefont {J.-L.}\ \bibnamefont {Wu}}, \bibinfo {author} {\bibfnamefont {Y.}~\bibnamefont {Wang}}, \bibinfo {author} {\bibfnamefont {J.-X.}\ \bibnamefont {Han}}, \bibinfo {author} {\bibfnamefont {Y.}~\bibnamefont {Jiang}}, \bibinfo {author} {\bibfnamefont {J.}~\bibnamefont {Song}}, \bibinfo {author} {\bibfnamefont {Y.}~\bibnamefont {Xia}}, \bibinfo {author} {\bibfnamefont {S.-L.}\ \bibnamefont {Su}},\ and\ \bibinfo {author} {\bibfnamefont {W.}~\bibnamefont {Li}},\ }\bibfield  {title} {\bibinfo {title} {Systematic-{{Error-Tolerant Multiqubit Holonomic Entangling Gates}}},\ }\href {https://doi.org/10.1103/PhysRevApplied.16.064031} {\bibfield  {journal} {\bibinfo  {journal} {Phys. Rev. Applied}\ }\textbf {\bibinfo {volume} {16}},\ \bibinfo {pages} {064031} (\bibinfo {year} {2021})}\BibitemShut {NoStop}%
\bibitem [{\citenamefont {Fan}\ \emph {et~al.}(2023)\citenamefont {Fan}, \citenamefont {Zhang}, \citenamefont {Jiao}, \citenamefont {Li}, \citenamefont {Bai}, \citenamefont {Wu}, \citenamefont {Zhao},\ and\ \citenamefont {Jia}}]{fan2023ManipulationSingleStoredphoton}%
  \BibitemOpen
  \bibfield  {author} {\bibinfo {author} {\bibfnamefont {J.}~\bibnamefont {Fan}}, \bibinfo {author} {\bibfnamefont {H.}~\bibnamefont {Zhang}}, \bibinfo {author} {\bibfnamefont {Y.}~\bibnamefont {Jiao}}, \bibinfo {author} {\bibfnamefont {C.}~\bibnamefont {Li}}, \bibinfo {author} {\bibfnamefont {J.}~\bibnamefont {Bai}}, \bibinfo {author} {\bibfnamefont {J.}~\bibnamefont {Wu}}, \bibinfo {author} {\bibfnamefont {J.}~\bibnamefont {Zhao}},\ and\ \bibinfo {author} {\bibfnamefont {S.}~\bibnamefont {Jia}},\ }\bibfield  {title} {\bibinfo {title} {Manipulation of single stored-photon with microwave field based on {{Rydberg}} polariton},\ }\href {https://doi.org/10.1364/OE.487471} {\bibfield  {journal} {\bibinfo  {journal} {Opt. Express}\ }\textbf {\bibinfo {volume} {31}},\ \bibinfo {pages} {20641} (\bibinfo {year} {2023})}\BibitemShut {NoStop}%
\bibitem [{\citenamefont {Dudin}\ and\ \citenamefont {Kuzmich}(2012)}]{dudin2012StronglyInteractingRydberg}%
  \BibitemOpen
  \bibfield  {author} {\bibinfo {author} {\bibfnamefont {Y.~O.}\ \bibnamefont {Dudin}}\ and\ \bibinfo {author} {\bibfnamefont {A.}~\bibnamefont {Kuzmich}},\ }\bibfield  {title} {\bibinfo {title} {Strongly {{Interacting Rydberg Excitations}} of a {{Cold Atomic Gas}}},\ }\href {https://doi.org/10.1126/science.1217901} {\bibfield  {journal} {\bibinfo  {journal} {Science}\ }\textbf {\bibinfo {volume} {336}},\ \bibinfo {pages} {887} (\bibinfo {year} {2012})}\BibitemShut {NoStop}%
\bibitem [{\citenamefont {Barredo}\ \emph {et~al.}(2015)\citenamefont {Barredo}, \citenamefont {Labuhn}, \citenamefont {Ravets}, \citenamefont {Lahaye}, \citenamefont {Browaeys},\ and\ \citenamefont {Adams}}]{Coherent_Barredo_2015}%
  \BibitemOpen
  \bibfield  {author} {\bibinfo {author} {\bibfnamefont {D.}~\bibnamefont {Barredo}}, \bibinfo {author} {\bibfnamefont {H.}~\bibnamefont {Labuhn}}, \bibinfo {author} {\bibfnamefont {S.}~\bibnamefont {Ravets}}, \bibinfo {author} {\bibfnamefont {T.}~\bibnamefont {Lahaye}}, \bibinfo {author} {\bibfnamefont {A.}~\bibnamefont {Browaeys}},\ and\ \bibinfo {author} {\bibfnamefont {C.~S.}\ \bibnamefont {Adams}},\ }\bibfield  {title} {\bibinfo {title} {Coherent excitation transfer in a spin chain of three rydberg atoms},\ }\href {https://doi.org/10.1103/PhysRevLett.114.113002} {\bibfield  {journal} {\bibinfo  {journal} {Phys. Rev. Lett.}\ }\textbf {\bibinfo {volume} {114}},\ \bibinfo {pages} {113002} (\bibinfo {year} {2015})}\BibitemShut {NoStop}%
\bibitem [{\citenamefont {Spong}\ \emph {et~al.}(2021)\citenamefont {Spong}, \citenamefont {Jiao}, \citenamefont {Hughes}, \citenamefont {Weatherill}, \citenamefont {Lesanovsky},\ and\ \citenamefont {Adams}}]{spong2021collectively}%
  \BibitemOpen
  \bibfield  {author} {\bibinfo {author} {\bibfnamefont {N.~L.~R.}\ \bibnamefont {Spong}}, \bibinfo {author} {\bibfnamefont {Y.}~\bibnamefont {Jiao}}, \bibinfo {author} {\bibfnamefont {O.~D.~W.}\ \bibnamefont {Hughes}}, \bibinfo {author} {\bibfnamefont {K.~J.}\ \bibnamefont {Weatherill}}, \bibinfo {author} {\bibfnamefont {I.}~\bibnamefont {Lesanovsky}},\ and\ \bibinfo {author} {\bibfnamefont {C.~S.}\ \bibnamefont {Adams}},\ }\bibfield  {title} {\bibinfo {title} {Collectively {{Encoded Rydberg Qubit}}},\ }\href {https://doi.org/10.1103/PhysRevLett.127.063604} {\bibfield  {journal} {\bibinfo  {journal} {Phys. Rev. Lett.}\ }\textbf {\bibinfo {volume} {127}},\ \bibinfo {pages} {063604} (\bibinfo {year} {2021})}\BibitemShut {NoStop}%
\bibitem [{\citenamefont {Festenstein}(2023)}]{festenstein2023arbitrary}%
  \BibitemOpen
  \bibfield  {author} {\bibinfo {author} {\bibfnamefont {M.}~\bibnamefont {Festenstein}},\ }\bibfield  {title} {\bibinfo {title} {An intuitive visualisation method for arbitrary qutrit (three level) states},\ }\href@noop {} {\bibfield  {journal} {\bibinfo  {journal} {arXiv preprint arXiv:2304.01741}\ } (\bibinfo {year} {2023})}\BibitemShut {NoStop}%
\bibitem [{\citenamefont {Pritchard}\ \emph {et~al.}(2010)\citenamefont {Pritchard}, \citenamefont {Maxwell}, \citenamefont {Gauguet}, \citenamefont {Weatherill}, \citenamefont {Jones},\ and\ \citenamefont {Adams}}]{pritchardCooperativeAtomLightInteraction2010}%
  \BibitemOpen
  \bibfield  {author} {\bibinfo {author} {\bibfnamefont {J.~D.}\ \bibnamefont {Pritchard}}, \bibinfo {author} {\bibfnamefont {D.}~\bibnamefont {Maxwell}}, \bibinfo {author} {\bibfnamefont {A.}~\bibnamefont {Gauguet}}, \bibinfo {author} {\bibfnamefont {K.~J.}\ \bibnamefont {Weatherill}}, \bibinfo {author} {\bibfnamefont {M.~P.~A.}\ \bibnamefont {Jones}},\ and\ \bibinfo {author} {\bibfnamefont {C.~S.}\ \bibnamefont {Adams}},\ }\bibfield  {title} {\bibinfo {title} {Cooperative {{Atom-Light Interaction}} in a {{Blockaded Rydberg Ensemble}}},\ }\href {https://doi.org/10.1103/PhysRevLett.105.193603} {\bibfield  {journal} {\bibinfo  {journal} {Phys. Rev. Lett.}\ }\textbf {\bibinfo {volume} {105}},\ \bibinfo {pages} {193603} (\bibinfo {year} {2010})}\BibitemShut {NoStop}%
\bibitem [{\citenamefont {Petrosyan}\ \emph {et~al.}(2011)\citenamefont {Petrosyan}, \citenamefont {Otterbach},\ and\ \citenamefont {Fleischhauer}}]{petrosyanElectromagneticallyInducedTransparency2011}%
  \BibitemOpen
  \bibfield  {author} {\bibinfo {author} {\bibfnamefont {D.}~\bibnamefont {Petrosyan}}, \bibinfo {author} {\bibfnamefont {J.}~\bibnamefont {Otterbach}},\ and\ \bibinfo {author} {\bibfnamefont {M.}~\bibnamefont {Fleischhauer}},\ }\bibfield  {title} {\bibinfo {title} {Electromagnetically {{Induced Transparency}} with {{Rydberg Atoms}}},\ }\href {https://doi.org/10.1103/PhysRevLett.107.213601} {\bibfield  {journal} {\bibinfo  {journal} {Phys. Rev. Lett.}\ }\textbf {\bibinfo {volume} {107}},\ \bibinfo {pages} {213601} (\bibinfo {year} {2011})}\BibitemShut {NoStop}%
\bibitem [{\citenamefont {Gorshkov}\ \emph {et~al.}(2011)\citenamefont {Gorshkov}, \citenamefont {Otterbach}, \citenamefont {Fleischhauer}, \citenamefont {Pohl},\ and\ \citenamefont {Lukin}}]{gorshkovPhotonPhotonInteractionsRydberg2011}%
  \BibitemOpen
  \bibfield  {author} {\bibinfo {author} {\bibfnamefont {A.~V.}\ \bibnamefont {Gorshkov}}, \bibinfo {author} {\bibfnamefont {J.}~\bibnamefont {Otterbach}}, \bibinfo {author} {\bibfnamefont {M.}~\bibnamefont {Fleischhauer}}, \bibinfo {author} {\bibfnamefont {T.}~\bibnamefont {Pohl}},\ and\ \bibinfo {author} {\bibfnamefont {M.~D.}\ \bibnamefont {Lukin}},\ }\bibfield  {title} {\bibinfo {title} {Photon-{{Photon Interactions}} via {{Rydberg Blockade}}},\ }\href {https://doi.org/10.1103/PhysRevLett.107.133602} {\bibfield  {journal} {\bibinfo  {journal} {Phys. Rev. Lett.}\ }\textbf {\bibinfo {volume} {107}},\ \bibinfo {pages} {133602} (\bibinfo {year} {2011})}\BibitemShut {NoStop}%
\bibitem [{\citenamefont {Maxwell}\ \emph {et~al.}(2013{\natexlab{a}})\citenamefont {Maxwell}, \citenamefont {Szwer}, \citenamefont {{Paredes-Barato}}, \citenamefont {Busche}, \citenamefont {Pritchard}, \citenamefont {Gauguet}, \citenamefont {Weatherill}, \citenamefont {Jones},\ and\ \citenamefont {Adams}}]{maxwellStorageControlOptical2013}%
  \BibitemOpen
  \bibfield  {author} {\bibinfo {author} {\bibfnamefont {D.}~\bibnamefont {Maxwell}}, \bibinfo {author} {\bibfnamefont {D.~J.}\ \bibnamefont {Szwer}}, \bibinfo {author} {\bibfnamefont {D.}~\bibnamefont {{Paredes-Barato}}}, \bibinfo {author} {\bibfnamefont {H.}~\bibnamefont {Busche}}, \bibinfo {author} {\bibfnamefont {J.~D.}\ \bibnamefont {Pritchard}}, \bibinfo {author} {\bibfnamefont {A.}~\bibnamefont {Gauguet}}, \bibinfo {author} {\bibfnamefont {K.~J.}\ \bibnamefont {Weatherill}}, \bibinfo {author} {\bibfnamefont {M.~P.~A.}\ \bibnamefont {Jones}},\ and\ \bibinfo {author} {\bibfnamefont {C.~S.}\ \bibnamefont {Adams}},\ }\bibfield  {title} {\bibinfo {title} {Storage and {{Control}} of {{Optical Photons Using Rydberg Polaritons}}},\ }\href {https://doi.org/10.1103/PhysRevLett.110.103001} {\bibfield  {journal} {\bibinfo  {journal} {Phys. Rev. Lett.}\ }\textbf {\bibinfo {volume} {110}},\ \bibinfo {pages} {103001} (\bibinfo {year} {2013}{\natexlab{a}})}\BibitemShut {NoStop}%
\bibitem [{\citenamefont {Jiao}\ \emph {et~al.}(2020)\citenamefont {Jiao}, \citenamefont {Spong}, \citenamefont {Hughes}, \citenamefont {So}, \citenamefont {Ilieva}, \citenamefont {Weatherill},\ and\ \citenamefont {Adams}}]{Jiao2020}%
  \BibitemOpen
  \bibfield  {author} {\bibinfo {author} {\bibfnamefont {Y.}~\bibnamefont {Jiao}}, \bibinfo {author} {\bibfnamefont {N.~L.~R.}\ \bibnamefont {Spong}}, \bibinfo {author} {\bibfnamefont {O.~D.~W.}\ \bibnamefont {Hughes}}, \bibinfo {author} {\bibfnamefont {C.}~\bibnamefont {So}}, \bibinfo {author} {\bibfnamefont {T.}~\bibnamefont {Ilieva}}, \bibinfo {author} {\bibfnamefont {K.~J.}\ \bibnamefont {Weatherill}},\ and\ \bibinfo {author} {\bibfnamefont {C.~S.}\ \bibnamefont {Adams}},\ }\bibfield  {title} {\bibinfo {title} {{Single-photon stored-light Ramsey interferometry using Rydberg polaritons}},\ }\href {https://doi.org/10.1364/OL.405143} {\bibfield  {journal} {\bibinfo  {journal} {Opt. Lett.}\ }\textbf {\bibinfo {volume} {45}},\ \bibinfo {pages} {5888} (\bibinfo {year} {2020})}\BibitemShut {NoStop}%
\bibitem [{\citenamefont {Otterbach}\ \emph {et~al.}(2013)\citenamefont {Otterbach}, \citenamefont {Moos}, \citenamefont {Muth},\ and\ \citenamefont {Fleischhauer}}]{otterbachWignerCrystallizationSingle2013}%
  \BibitemOpen
  \bibfield  {author} {\bibinfo {author} {\bibfnamefont {J.}~\bibnamefont {Otterbach}}, \bibinfo {author} {\bibfnamefont {M.}~\bibnamefont {Moos}}, \bibinfo {author} {\bibfnamefont {D.}~\bibnamefont {Muth}},\ and\ \bibinfo {author} {\bibfnamefont {M.}~\bibnamefont {Fleischhauer}},\ }\bibfield  {title} {\bibinfo {title} {Wigner {{Crystallization}} of {{Single Photons}} in {{Cold Rydberg Ensembles}}},\ }\href {https://doi.org/10.1103/PhysRevLett.111.113001} {\bibfield  {journal} {\bibinfo  {journal} {Phys. Rev. Lett.}\ }\textbf {\bibinfo {volume} {111}},\ \bibinfo {pages} {113001} (\bibinfo {year} {2013})}\BibitemShut {NoStop}%
\bibitem [{\citenamefont {Gullans}\ \emph {et~al.}(2016)\citenamefont {Gullans}, \citenamefont {Thompson}, \citenamefont {Wang}, \citenamefont {Liang}, \citenamefont {Vuleti{\'c}}, \citenamefont {Lukin},\ and\ \citenamefont {Gorshkov}}]{gullansEffectiveFieldTheory2016}%
  \BibitemOpen
  \bibfield  {author} {\bibinfo {author} {\bibfnamefont {M.~J.}\ \bibnamefont {Gullans}}, \bibinfo {author} {\bibfnamefont {J.~D.}\ \bibnamefont {Thompson}}, \bibinfo {author} {\bibfnamefont {Y.}~\bibnamefont {Wang}}, \bibinfo {author} {\bibfnamefont {Q.-Y.}\ \bibnamefont {Liang}}, \bibinfo {author} {\bibfnamefont {V.}~\bibnamefont {Vuleti{\'c}}}, \bibinfo {author} {\bibfnamefont {M.~D.}\ \bibnamefont {Lukin}},\ and\ \bibinfo {author} {\bibfnamefont {A.~V.}\ \bibnamefont {Gorshkov}},\ }\bibfield  {title} {\bibinfo {title} {Effective {{Field Theory}} for {{Rydberg Polaritons}}},\ }\href {https://doi.org/10.1103/PhysRevLett.117.113601} {\bibfield  {journal} {\bibinfo  {journal} {Phys. Rev. Lett.}\ }\textbf {\bibinfo {volume} {117}},\ \bibinfo {pages} {113601} (\bibinfo {year} {2016})}\BibitemShut {NoStop}%
\bibitem [{\citenamefont {Browaeys}\ and\ \citenamefont {Lahaye}(2020)}]{browaeysManybodyPhysicsIndividually2020}%
  \BibitemOpen
  \bibfield  {author} {\bibinfo {author} {\bibfnamefont {A.}~\bibnamefont {Browaeys}}\ and\ \bibinfo {author} {\bibfnamefont {T.}~\bibnamefont {Lahaye}},\ }\bibfield  {title} {\bibinfo {title} {Many-body physics with individually controlled {{Rydberg}} atoms},\ }\href {https://doi.org/10.1038/s41567-019-0733-z} {\bibfield  {journal} {\bibinfo  {journal} {Nat. Phys.}\ }\textbf {\bibinfo {volume} {16}},\ \bibinfo {pages} {132} (\bibinfo {year} {2020})}\BibitemShut {NoStop}%
\bibitem [{\citenamefont {Fleischhauer}\ and\ \citenamefont {Lukin}(2000)}]{Fleischhauer2000}%
  \BibitemOpen
  \bibfield  {author} {\bibinfo {author} {\bibfnamefont {M.}~\bibnamefont {Fleischhauer}}\ and\ \bibinfo {author} {\bibfnamefont {M.~D.}\ \bibnamefont {Lukin}},\ }\bibfield  {title} {\bibinfo {title} {{Dark-state polaritons in electromagnetically induced transparency}},\ }\href {https://doi.org/10.1103/PhysRevLett.84.5094} {\bibfield  {journal} {\bibinfo  {journal} {Phys. Rev. Lett.}\ }\textbf {\bibinfo {volume} {84}},\ \bibinfo {pages} {5094} (\bibinfo {year} {2000})}\BibitemShut {NoStop}%
\bibitem [{\citenamefont {Maxwell}\ \emph {et~al.}(2013{\natexlab{b}})\citenamefont {Maxwell}, \citenamefont {Szwer}, \citenamefont {Paredes-Barato}, \citenamefont {Busche}, \citenamefont {Pritchard}, \citenamefont {Gauguet}, \citenamefont {Weatherill}, \citenamefont {Jones},\ and\ \citenamefont {Adams}}]{Storage_Maxwell_2013}%
  \BibitemOpen
  \bibfield  {author} {\bibinfo {author} {\bibfnamefont {D.}~\bibnamefont {Maxwell}}, \bibinfo {author} {\bibfnamefont {D.~J.}\ \bibnamefont {Szwer}}, \bibinfo {author} {\bibfnamefont {D.}~\bibnamefont {Paredes-Barato}}, \bibinfo {author} {\bibfnamefont {H.}~\bibnamefont {Busche}}, \bibinfo {author} {\bibfnamefont {J.~D.}\ \bibnamefont {Pritchard}}, \bibinfo {author} {\bibfnamefont {A.}~\bibnamefont {Gauguet}}, \bibinfo {author} {\bibfnamefont {K.~J.}\ \bibnamefont {Weatherill}}, \bibinfo {author} {\bibfnamefont {M.~P.~A.}\ \bibnamefont {Jones}},\ and\ \bibinfo {author} {\bibfnamefont {C.~S.}\ \bibnamefont {Adams}},\ }\bibfield  {title} {\bibinfo {title} {Storage and control of optical photons using rydberg polaritons},\ }\href {https://doi.org/10.1103/PhysRevLett.110.103001} {\bibfield  {journal} {\bibinfo  {journal} {Phys. Rev. Lett.}\ }\textbf {\bibinfo {volume} {110}},\ \bibinfo {pages} {103001} (\bibinfo {year} {2013}{\natexlab{b}})}\BibitemShut {NoStop}%
\bibitem [{\citenamefont {Dudin}\ \emph {et~al.}(2012)\citenamefont {Dudin}, \citenamefont {Li}, \citenamefont {Bariani},\ and\ \citenamefont {Kuzmich}}]{Dudin2012}%
  \BibitemOpen
  \bibfield  {author} {\bibinfo {author} {\bibfnamefont {Y.~O.}\ \bibnamefont {Dudin}}, \bibinfo {author} {\bibfnamefont {L.}~\bibnamefont {Li}}, \bibinfo {author} {\bibfnamefont {F.}~\bibnamefont {Bariani}},\ and\ \bibinfo {author} {\bibfnamefont {A.}~\bibnamefont {Kuzmich}},\ }\bibfield  {title} {\bibinfo {title} {{Observation of coherent many-body Rabi oscillations}},\ }\href {https://doi.org/10.1038/nphys2413} {\bibfield  {journal} {\bibinfo  {journal} {Nat. Phys.}\ }\textbf {\bibinfo {volume} {8}},\ \bibinfo {pages} {790} (\bibinfo {year} {2012})}\BibitemShut {NoStop}%
\bibitem [{\citenamefont {Ornelas-Huerta}\ \emph {et~al.}(2020)\citenamefont {Ornelas-Huerta}, \citenamefont {Craddock}, \citenamefont {Goldschmidt}, \citenamefont {Hachtel}, \citenamefont {Wang}, \citenamefont {Bienias}, \citenamefont {Gorshkov}, \citenamefont {Rolston},\ and\ \citenamefont {Porto}}]{Ornelas-Huerta2020}%
  \BibitemOpen
  \bibfield  {author} {\bibinfo {author} {\bibfnamefont {D.~P.}\ \bibnamefont {Ornelas-Huerta}}, \bibinfo {author} {\bibfnamefont {A.~N.}\ \bibnamefont {Craddock}}, \bibinfo {author} {\bibfnamefont {E.~A.}\ \bibnamefont {Goldschmidt}}, \bibinfo {author} {\bibfnamefont {A.~J.}\ \bibnamefont {Hachtel}}, \bibinfo {author} {\bibfnamefont {Y.}~\bibnamefont {Wang}}, \bibinfo {author} {\bibfnamefont {P.}~\bibnamefont {Bienias}}, \bibinfo {author} {\bibfnamefont {A.~V.}\ \bibnamefont {Gorshkov}}, \bibinfo {author} {\bibfnamefont {S.~L.}\ \bibnamefont {Rolston}},\ and\ \bibinfo {author} {\bibfnamefont {J.~V.}\ \bibnamefont {Porto}},\ }\bibfield  {title} {\bibinfo {title} {{On-demand indistinguishable single photons from an efficient and pure source based on a Rydberg ensemble}},\ }\href {https://doi.org/10.1364/optica.391485} {\bibfield  {journal} {\bibinfo  {journal} {Optica}\ }\textbf {\bibinfo {volume} {7}},\ \bibinfo {pages} {813} (\bibinfo {year} {2020})}\BibitemShut {NoStop}%
\bibitem [{\citenamefont {Beterov}\ \emph {et~al.}(2009)\citenamefont {Beterov}, \citenamefont {Ryabtsev}, \citenamefont {Tretyakov},\ and\ \citenamefont {Entin}}]{beterov2009}%
  \BibitemOpen
  \bibfield  {author} {\bibinfo {author} {\bibfnamefont {I.~I.}\ \bibnamefont {Beterov}}, \bibinfo {author} {\bibfnamefont {I.~I.}\ \bibnamefont {Ryabtsev}}, \bibinfo {author} {\bibfnamefont {D.~B.}\ \bibnamefont {Tretyakov}},\ and\ \bibinfo {author} {\bibfnamefont {V.~M.}\ \bibnamefont {Entin}},\ }\bibfield  {title} {\bibinfo {title} {Quasiclassical calculations of blackbody-radiation-induced depopulation rates and effective lifetimes of {{Rydberg}} n {{S}} , n {{P}} , and n {{D}} alkali-metal atoms with n {$\leq$} 80},\ }\href {https://doi.org/10.1103/PhysRevA.79.052504} {\bibfield  {journal} {\bibinfo  {journal} {Physical Review A}\ }\textbf {\bibinfo {volume} {79}},\ \bibinfo {pages} {052504} (\bibinfo {year} {2009})}\BibitemShut {NoStop}%
\bibitem [{\citenamefont {Zeiher}\ \emph {et~al.}(2016)\citenamefont {Zeiher}, \citenamefont {Van~Bijnen}, \citenamefont {Schau{\ss}}, \citenamefont {Hild}, \citenamefont {Choi}, \citenamefont {Pohl}, \citenamefont {Bloch},\ and\ \citenamefont {Gross}}]{zeiher2016}%
  \BibitemOpen
  \bibfield  {author} {\bibinfo {author} {\bibfnamefont {J.}~\bibnamefont {Zeiher}}, \bibinfo {author} {\bibfnamefont {R.}~\bibnamefont {Van~Bijnen}}, \bibinfo {author} {\bibfnamefont {P.}~\bibnamefont {Schau{\ss}}}, \bibinfo {author} {\bibfnamefont {S.}~\bibnamefont {Hild}}, \bibinfo {author} {\bibfnamefont {J.-y.}\ \bibnamefont {Choi}}, \bibinfo {author} {\bibfnamefont {T.}~\bibnamefont {Pohl}}, \bibinfo {author} {\bibfnamefont {I.}~\bibnamefont {Bloch}},\ and\ \bibinfo {author} {\bibfnamefont {C.}~\bibnamefont {Gross}},\ }\bibfield  {title} {\bibinfo {title} {Many-body interferometry of a {{Rydberg-dressed}} spin lattice},\ }\href {https://doi.org/10.1038/nphys3835} {\bibfield  {journal} {\bibinfo  {journal} {Nature Physics}\ }\textbf {\bibinfo {volume} {12}},\ \bibinfo {pages} {1095} (\bibinfo {year} {2016})}\BibitemShut {NoStop}%
\bibitem [{\citenamefont {De~L{\'e}s{\'e}leuc}\ \emph {et~al.}(2018)\citenamefont {De~L{\'e}s{\'e}leuc}, \citenamefont {Barredo}, \citenamefont {Lienhard}, \citenamefont {Browaeys},\ and\ \citenamefont {Lahaye}}]{deleseleuc2018}%
  \BibitemOpen
  \bibfield  {author} {\bibinfo {author} {\bibfnamefont {S.}~\bibnamefont {De~L{\'e}s{\'e}leuc}}, \bibinfo {author} {\bibfnamefont {D.}~\bibnamefont {Barredo}}, \bibinfo {author} {\bibfnamefont {V.}~\bibnamefont {Lienhard}}, \bibinfo {author} {\bibfnamefont {A.}~\bibnamefont {Browaeys}},\ and\ \bibinfo {author} {\bibfnamefont {T.}~\bibnamefont {Lahaye}},\ }\bibfield  {title} {\bibinfo {title} {Analysis of imperfections in the coherent optical excitation of single atoms to {{Rydberg}} states},\ }\href {https://doi.org/10.1103/PhysRevA.97.053803} {\bibfield  {journal} {\bibinfo  {journal} {Physical Review A}\ }\textbf {\bibinfo {volume} {97}},\ \bibinfo {pages} {053803} (\bibinfo {year} {2018})}\BibitemShut {NoStop}%
\bibitem [{\citenamefont {Tan}\ \emph {et~al.}(2018)\citenamefont {Tan}, \citenamefont {Zhang}, \citenamefont {Liu}, \citenamefont {Xue}, \citenamefont {Yu}, \citenamefont {Zhu}, \citenamefont {Yan}, \citenamefont {Zhu},\ and\ \citenamefont {Yu}}]{tanTopologicalMaxwellMetal2018}%
  \BibitemOpen
  \bibfield  {author} {\bibinfo {author} {\bibfnamefont {X.}~\bibnamefont {Tan}}, \bibinfo {author} {\bibfnamefont {D.-W.}\ \bibnamefont {Zhang}}, \bibinfo {author} {\bibfnamefont {Q.}~\bibnamefont {Liu}}, \bibinfo {author} {\bibfnamefont {G.}~\bibnamefont {Xue}}, \bibinfo {author} {\bibfnamefont {H.-F.}\ \bibnamefont {Yu}}, \bibinfo {author} {\bibfnamefont {Y.-Q.}\ \bibnamefont {Zhu}}, \bibinfo {author} {\bibfnamefont {H.}~\bibnamefont {Yan}}, \bibinfo {author} {\bibfnamefont {S.-L.}\ \bibnamefont {Zhu}},\ and\ \bibinfo {author} {\bibfnamefont {Y.}~\bibnamefont {Yu}},\ }\bibfield  {title} {\bibinfo {title} {Topological {{Maxwell Metal Bands}} in a {{Superconducting Qutrit}}},\ }\href@noop {} {\bibfield  {journal} {\bibinfo  {journal} {PHYSICAL REVIEW LETTERS}\ } (\bibinfo {year} {2018})}\BibitemShut {NoStop}%
\bibitem [{\citenamefont {{Paredes-Barato}}\ and\ \citenamefont {Adams}(2014)}]{paredes-barato2014AllOpticalQuantumInformationb}%
  \BibitemOpen
  \bibfield  {author} {\bibinfo {author} {\bibfnamefont {D.}~\bibnamefont {{Paredes-Barato}}}\ and\ \bibinfo {author} {\bibfnamefont {C.~S.}\ \bibnamefont {Adams}},\ }\bibfield  {title} {\bibinfo {title} {All-{{Optical Quantum Information Processing Using Rydberg Gates}}},\ }\href {https://doi.org/10.1103/PhysRevLett.112.040501} {\bibfield  {journal} {\bibinfo  {journal} {Physical Review Letters}\ }\textbf {\bibinfo {volume} {112}},\ \bibinfo {pages} {040501} (\bibinfo {year} {2014})}\BibitemShut {NoStop}%
\bibitem [{\citenamefont {Busche}\ \emph {et~al.}(2017)\citenamefont {Busche}, \citenamefont {Huillery}, \citenamefont {Ball}, \citenamefont {Ilieva}, \citenamefont {Jones},\ and\ \citenamefont {Adams}}]{Busche2017}%
  \BibitemOpen
  \bibfield  {author} {\bibinfo {author} {\bibfnamefont {H.}~\bibnamefont {Busche}}, \bibinfo {author} {\bibfnamefont {P.}~\bibnamefont {Huillery}}, \bibinfo {author} {\bibfnamefont {S.~W.}\ \bibnamefont {Ball}}, \bibinfo {author} {\bibfnamefont {T.}~\bibnamefont {Ilieva}}, \bibinfo {author} {\bibfnamefont {M.~P.}\ \bibnamefont {Jones}},\ and\ \bibinfo {author} {\bibfnamefont {C.~S.}\ \bibnamefont {Adams}},\ }\bibfield  {title} {\bibinfo {title} {{Contactless nonlinear optics mediated by long-range Rydberg interactions}},\ }\href {https://doi.org/10.1038/nphys4058} {\bibfield  {journal} {\bibinfo  {journal} {Nat. Phys.}\ }\textbf {\bibinfo {volume} {13}},\ \bibinfo {pages} {655} (\bibinfo {year} {2017})}\BibitemShut {NoStop}%
\bibitem [{\citenamefont {Khazali}\ \emph {et~al.}(2019)\citenamefont {Khazali}, \citenamefont {Murray},\ and\ \citenamefont {Pohl}}]{khazali2019PolaritonExchangeInteractions}%
  \BibitemOpen
  \bibfield  {author} {\bibinfo {author} {\bibfnamefont {M.}~\bibnamefont {Khazali}}, \bibinfo {author} {\bibfnamefont {C.~R.}\ \bibnamefont {Murray}},\ and\ \bibinfo {author} {\bibfnamefont {T.}~\bibnamefont {Pohl}},\ }\bibfield  {title} {\bibinfo {title} {Polariton {{Exchange Interactions}} in {{Multichannel Optical Networks}}},\ }\href {https://doi.org/10.1103/PhysRevLett.123.113605} {\bibfield  {journal} {\bibinfo  {journal} {Physical Review Letters}\ }\textbf {\bibinfo {volume} {123}},\ \bibinfo {pages} {113605} (\bibinfo {year} {2019})}\BibitemShut {NoStop}%
\bibitem [{\citenamefont {Yang}\ \emph {et~al.}(2024)\citenamefont {Yang}, \citenamefont {Luo}, \citenamefont {Zhang}, \citenamefont {Wang}, \citenamefont {Zou}, \citenamefont {Xia},\ and\ \citenamefont {Lu}}]{yang2024a}%
  \BibitemOpen
  \bibfield  {author} {\bibinfo {author} {\bibfnamefont {Y.~A.}\ \bibnamefont {Yang}}, \bibinfo {author} {\bibfnamefont {W.-T.}\ \bibnamefont {Luo}}, \bibinfo {author} {\bibfnamefont {J.-L.}\ \bibnamefont {Zhang}}, \bibinfo {author} {\bibfnamefont {S.-Z.}\ \bibnamefont {Wang}}, \bibinfo {author} {\bibfnamefont {C.-L.}\ \bibnamefont {Zou}}, \bibinfo {author} {\bibfnamefont {T.}~\bibnamefont {Xia}},\ and\ \bibinfo {author} {\bibfnamefont {Z.-T.}\ \bibnamefont {Lu}},\ }\bibfield  {title} {\bibinfo {title} {Minute-scale {{Schr{\"o}dinger-cat}} state of spin-5/2 atoms},\ }\href {https://doi.org/10.1038/s41566-024-01555-3} {\bibfield  {journal} {\bibinfo  {journal} {Nature Photonics}\ ,\ \bibinfo {pages} {1}} (\bibinfo {year} {2024})}\BibitemShut {NoStop}%
\end{thebibliography}%
\end{document}